\documentclass[twocolumn,showpacs,preprintnumbers,amsmath,amssymb]{revtex4}
\usepackage{amsfonts, amsmath, bm, bbm, graphicx, epsfig, epstopdf}
\usepackage{dcolumn}
\usepackage{textcomp}
\usepackage{soul}
\usepackage{graphicx}
\usepackage{amsfonts, amsmath, bm, bbm, graphicx, epsfig, epstopdf}
\usepackage{dcolumn}
\usepackage{textcomp}
 \usepackage[caption=false]{subfig}
\usepackage{caption}

\usepackage{soul}

\begin{document}

\bibliographystyle{apsrev}

\title{Fast multi-qubit gates by adiabatic evolution in interacting excited state manifolds}

\author{Mohammadsadegh Khazali$^{1,2}$ and Klaus M\o lmer$^3$}
\affiliation{$^1$ Department of Physics, Sharif University of Technology, Tehran 14588, Iran \\
$^2$ School of Nano Science, Institute for Research in Fundamental Sciences (IPM), Tehran 19395-5531, Iran\\
$^3$ Department of Physics and Astronomy, Aarhus University , DK 8000 Aarhus C, Denmark}

\date{\today}
\begin{abstract}
Quantum computing and quantum simulation can be implemented by concatenation of one- and two-qubit gates and interactions. For most physical implementations, however, it may be advantageous to explore state components and interactions that depart from this universal paradigm and offer faster or more robust access to more advanced operations on the system.
In this article, we show that adiabatic passage along the dark eigenstate of excitation exchange interactions can be used to implement fast multi-qubit Toffoli (C$_k$-NOT) and fan-out (C-NOT$^k$) gates. This mechanism can be realized by simultaneous excitation of atoms to Rydberg levels, featuring resonant exchange interaction.
Our theoretical estimates and numerical simulations show that these multi-qubit Rydberg gates are possible with errors below 1\% for up to 20 qubits.
The excitation exchange mechanism is ubiquitous across experimental platforms and we show that similar multi-qubit gates can be implemented in superconducting circuits.
\end{abstract}

\maketitle

\section{Introduction}
\label{intro}
In the circuit model paradigm of quantum computing, a quantum algorithm is implemented as a sequence of one- and two-qubit gates chosen from a suitable universal gate set \cite{Bar95}. This paradigm mimics the models of classical computers and it allows direct assessment of the potential of any candidate quantum system for quantum computing. Early proposals for quantum computing were thus largely based on the identification of qubit degrees of freedom in a physical system and interactions suitable for two-qubit gates. The circuit model permits comparison of the achievements of different candidate systems for quantum computing by the execution time and fidelity of their one- and two-qubit gates, and much effort has thus gone into minimization of the one- and two-qubit gate errors.

By concatenating infinitesimal steps of time evolution with the one and two-body gate interactions, it is possible to synthesize more complex effective interactions, \cite{Wec14,Bab14,Pou15}, but it has also been realized that the physical properties of  many quantum information candidate systems already yield effective  many-body interactions and allow efficient implementation of certain multi-qubit gates \cite{Mar16,Ise11}. Use of system specific properties may significantly reduce the number of operations and hence the errors incurred during execution of a quantum algorithm. It complicates the transfer of ideas and comparison of performance among candidate platforms for quantum computing and simulation, and along with the quantitative assessment of the performance of a given scheme, it is hence worthwhile to identify and emphasize its generic features and properties.

{\it Trapped ions.} Quantum computing with trapped ions employs the coupling of electronic excitation degrees of freedom to a common motional vibration, and permits a direct implementation of the Toffoli gate \cite{Mon09}. Rather than the original two-qubit gates \cite{Cir95,Mol99}, experimentalists now routinely employ the joint vibrational coupling of all ions to effectuate global operations on the entire quantum register. A global $S_x^2$ collective spin interaction, mediated by a single vibrational mode can produce GHZ states in a single laser pulse \cite{Sor00,Mol99b,Sac00,Lei05,Bla08,Mon11}, and the multi-qubit $S_x^2$ interaction is sufficient together with single qubit phase gates for universal quantum computing \cite{Mol99,Mar16} and quantum simulation \cite{Lan11}. Ion qubit interactions mediated by multiple vibrational modes also permit physical motivated shortcuts to simulation of spin models with tailored finite range interactions \cite{Kor12}.

{\it Superconducting qubits.} Superconducting qubits are addressed and manipulated by their interaction with cavities or waveguides, and
multi-qubit operations on superconducting qubits in coupled cavities are discussed in \cite{Ral07,Fed12,Ree12}. A recent manuscript presents a scheme for single step implementation of multi-qubit Toffoli gates on superconducting architectures without cavities but with suitable Ising qubit-qubit interactions \cite{Ras19}. Quantum simulations of complex evolution dynamics benefit from the physical properties offered by the superconducting qubit systems and constitute an expanding field with considerable recent progress \cite{Sla15,Rou17,Lan17}.

{\it Atoms in optical lattices.} Ground state interactions of atoms in optical lattice or tweezer potentials have been proposed for neutral atom quantum gates \cite{Man03,Tre06,Jor14,Kau15,Les18}. Lattice displacements and tunnelling in conjunction with interactions may simultaneously affect many atoms and may hence be used for efficient quantum simulators of complex many-body spatial and spin dynamics, see for example \cite{Sor99,Tre06,Blo12,Blo18}   .

{\it Rydberg excited atoms.} Another prominent scheme for quantum computing and simulation with neutral atoms employs lasers to excite atoms to high lying and strongly interacting Rydberg states. This gives rise to the excitation blockade mechanism \cite{Jak00}, which also works on ensemble qubits \cite{Luk01}. The ability of one atom to block excitation of a whole surrounding ensemble has led to proposals for multi-qubit gates \cite{Ise11,Mol11,Wei10,Ber17,Kee19}. It has been proposed to use adiabatic passage processes to prepare strongly entangled states \cite{Una02} and to entangle a single atom with a mesoscopic ensemble \cite{Mul09}, and in this article we presents a new, robust adiabatic passage mechanism for multi-qubit gates. Rather than the blockade mechanism, our gate makes use of the strong dipolar exchange interactions between Rydberg excited atoms and employs adiabatic following of a multiply Rydberg excited dark eigenstate under variation of laser excitation amplitudes. This leads to robust implementation of the fan-out and Toffoli gates with infidelities at the 1\% level for up to $k=20$ control and target qubits. Such small  errors are compatible with surface error correction codes for quantum computing \cite{Wan11} while the multi-partite GHZ states, prepared by a single pulse, are of sufficient fidelity to offer metrological advances.

As the above mentioned proposals are all very specific to the given system, and they provide shortcuts of very different nature to multi-qubit operations, it is difficult to compare their performance. It may also be difficult to generally assess the precise gain of adding a very specific multi-qubit gate operation to the already available universal gate set.
We shall argue, however, that the multi-qubit fan-out and Toffoli gates are so versatile and so prominently applied in quantum computing \cite{Mol11, Van01}, simulation and error correction algorithms \cite{Nie11,Cor98}, that precisely these gates are worthwhile including and optimizing in any architecture where they can be implemented efficiently. While Rydberg gates still lag behind the fidelities of, e.g., the two-qubit ion trap gates, with the multi-qubit Rydberg gate capacity presented here, the neutral atom proposals may enter the stage and, for certain tasks, even supersede the ion performance and offer a more promising path for extension towards hundreds or thousands of bits. The Rydberg level structure and long range interactions are crucial for the detailed functioning of the gates, but the central idea of the present proposal, i.e., the adiabatic following of the eigenstates of a multi-qubit system subject to resonant exchange interactions may be implemented in other systems, as we shall demonstrate for superconducting circuits.

The article is organized as follows. In Sec.~\ref{sec:RydQI}, we briefly summarize recent experimental and theoretical progress with Rydberg atom quantum gates, and we describe the level scheme and excitation protocols of our multi-qubit Toffoli and fan-out gates.
In Sec.~\ref{sec_dark states} we present the derivation of the multi-atom dark state responsible for the functioning of the gates.
In Sec.~\ref{Sec_Error} we discuss error sources and estimate their impact on the quantum gate fidelities. In Sec. ~\ref{sec:sup}, we sketch how key elements of our protocol can be applied to superconducting qubits.
In the Appendices, we supplement our error estimates with more detailed models and we present numerical simulations in support of the error scalings identified in the main text.

\section{Rydberg atom quantum gates}
\label{sec:RydQI}

The strong dipolar interaction between Rydberg excited atoms has a high potential for application in quantum information processing \cite{Jak00,Luk01,Saf10,Ada19, Urb09,Gae09, Beg13,Mal15,Gra19,Lev19,Jau16,Kha15,Fri05, Gor11,Tia19,And16,Par15,Bin14,Bus17} and for implementation and simulation of quantum  many-body physics with neutral atoms \cite{Wei10,Lie18, Ber17,Kee19,Kha16,Kha18,Saf09}. Single qubit gates are achieved by coherent driving of transitions between the  ground hyperfine qubit states in the individual atoms, while in, e.g., the  two-qubit blockade gate \cite{Jak00}, Rydberg excitation of the control qubit atom shifts the nearby target atom Rydberg state energy enough to prevent its subsequent laser excitation.
After years of experimental research, we now have atomic arrays with tens or hundreds of atoms, and the proposal to use Rydberg interactions for neutral atom quantum computing has finally reached the conditions for high fidelity operations \cite{Gra19,Lev19}.

Our goal in this article is to propose and analyze multi-qubit gates, and the Rydberg blockade mechanism may, indeed, apply simultaneously to a number of target atoms which are all shifted in energy and which may thus undergo the same conditional evolution due to the excitation of a single control atom, as, e.g., required in the fan-out gate, implemented as a C-NOT$^k$ with one control and $k$ target qubits.
It is also possible to implement a multi-qubit Toffoli, C$_2$-NOT  \cite{Shi18,Bet18} C$_k$-NOT  \cite{Ise11} and  C$_k$-Z \cite{Mol11,Pet16}, conditioned on $k$ control qubits being all in the logical state $|1\rangle$.

Already for two atoms, however, the blockade scheme is vulnerable to weak off-resonant excitation, accumulation of phase errors and the dipolar forces cause decoherence through entanglement with the relative atomic motion \cite{Saf10}, and these errors accumulate proportional to the number of qubits in the proposed multi-qubit gate operations \cite{Ise11}. Recently a dark state approach was proposed to enhance the fidelity of Rydberg two-qubit gates \cite{Pet17}. In that proposal resonant laser excitation and dipole-dipole interaction between pairs of Rydberg product states result in the formation of an adiabatically varying energy eigenstate with vanishing energy shift. Adiabatic following of this ``two-atom dark state'' is inherently robust, it yields very little population and phase error, and it induces no forces between the atoms.

In the following sections, we provide schemes to implement the C-NOT$^k$ and C$_k$-NOT gates on an arbitrary number of atoms by similar adiabatic processes. It is a priori far from clear that a similar dark state would exist in a multi-atom system, where simultaneous excitation of a large number of atoms may occur, and the dipolar exchange interaction couples all atoms in a highly intricate manner. If one, optimistically, assumes that a single collectively and multiply excited dark state could be found, one might reasonably fear that any variation in atomic interaction strengths would break the symmetry of the state and cause population of state components in a highly non-controllable manner. But this does not happen! We show by analyses and simulations that a simple pulse sequence, indeed, drives the many-atom state along a multi-atom excited eigenstate and back to the stable qubit states, in a manner that implements the desired gates with up to 20 control or target qubits.

\subsection{Toffoli and fan-out gate schemes}
\label{sec:Scheme}

\begin{figure}
\centering
\scalebox{0.38}{\includegraphics*[viewport=0 200 700 500]{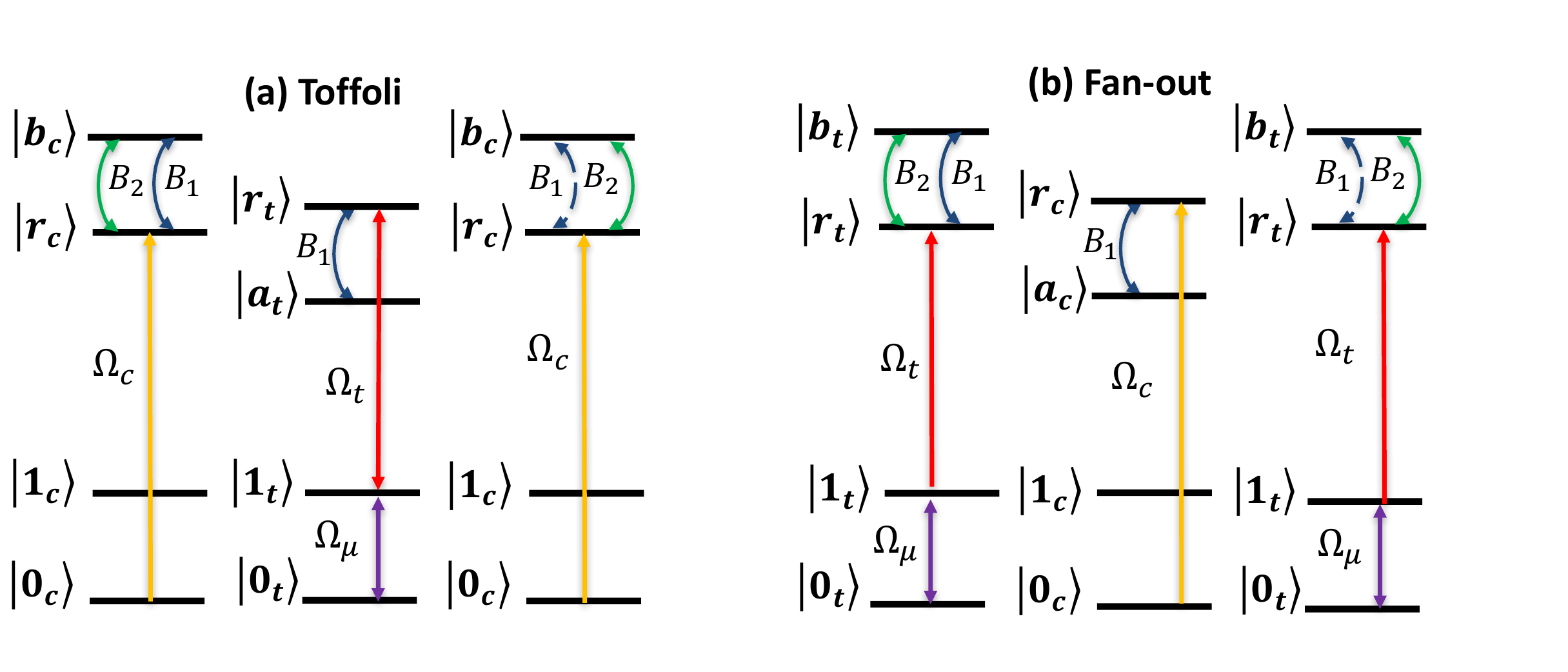}}
\caption{ Level scheme for three qubit  (a) Toffoli and  (b) fan-out gates.
States $|0_{c,t}\rangle$, $|1_{c,t}\rangle$ are long-lived qubit basis states and $|r_{c,t}\rangle$, $|a_{c,t}\rangle$ and $|b_{c,t}\rangle$ are Rydberg states of the control and target atoms.
Hadamard qubit gates ($\pi/2$-pulses) on the target atoms are implemented by the $\Omega_{\mu}$ classical field. Control states $|0_{c}\rangle$ and target states $|1_{t}\rangle$ states are coupled to the Rydberg levels  $|r_{c,t}\rangle$ by $\Omega_{c,t}$ lasers.
The resonant exchange interaction between control and target Rydberg atoms $|r_cr_t\rangle \leftrightarrow |a_{c,t}b_{c,t}\rangle$  with strength  $B_1\gg \Omega_t$ drives  the dark state dynamics, while  an intra component exchange interaction $B_2$ perturbs the gate operation. }\label{DarkStateToffoli}
\end{figure}

The physical implementation of our multi-qubit gates is illustrated in Fig.  \ref{DarkStateToffoli}.
The qubit basis consists of long lived hyperfine ground levels $|0\rangle$ and $|1\rangle$ that can be coupled coherently with Rabi frequency $\Omega_\mu$ with a microwave or optical Raman transition. The qubit states  $|0_{c}\rangle$ and $|1_{t}\rangle$ are resonantly coupled with Rabi frequency $\Omega_{c,t}$ to the Rydberg state $|r_{c,t}\rangle$ by a one-photon or two-photon process. We assume that a resonant dipolar interaction with strength $B_1$ couples near resonant product states $|r_{c},r_{t}\rangle \equiv |r_{c}\rangle\otimes|r_{t}\rangle$ and $|a_{c,t},b_{c,t}\rangle$, and we incorporate a resonant coupling with strength $B_2$ among the control or target product states $|b_{c,t},r_{c,t}\rangle$ and $|r_{c,t},b_{c,t}\rangle$.

{\bf Toffoli, C$_k$-NOT gate}
The Toffoli gate applies a NOT operation to the target qubit, conditioned on all control qubits being in state $|1_c\rangle$. We compose the Toffoli gate by application of target qubit Hadamard operations before and after a C$_k$-phase gate implemented with the following steps:
A simultaneous $\pi$ pulse transfers those control atoms that are in  $|0_c\rangle$ to the Rydberg state $|r_c\rangle$. By supplying external fields we can tune the atomic resonances so that these excited atoms experience negligible mutual interaction  for the chosen interatomic distances, see App.  A1, A4.
We then apply a smooth $2\pi$ pulse to the single target atom in resonance with the  $|1_t\rangle - |r_t\rangle$ transition. In the absence of any Rydberg excited control atom, the target state $|1_t\rangle$ thus acquires a factor $(-1)$, while the presence of any Rydberg excited control atom $|r_c\rangle$ leads to adiabatic following along a dark state with no phase shift (see next section).
Application of a second simultaneous  $\pi$  pulse on the control atoms returns the $|r_c\rangle$ population to  $|0_c\rangle$.

{\bf Fan-out,  C-NOT$^k$ gate}
The fan-out gate is equivalent to the application of a C-NOT gate with a single control qubit and a number of target qubits. It can be carried out sequentially, but we propose a protocol using simultaneous adiabatic driving of the joint state of the control and all the target qubits. Like the Toffoli gate, the fan-out gate  applies Hadamard target qubit gates before and after a suitable phase gate:
A $\pi$ pulse transfers  $|0_c\rangle$ to the control atom Rydberg state $|r_c\rangle$, followed by
a smooth $2\pi$ pulse on the   $|1_t\rangle - |r_t\rangle$ target atom transition. If the control atom is not Rydberg excited (populates $|1_c\rangle$), the target atoms evolve independently and acquire a phase by the $2\pi$ pulse, while in the presence of the $|r_c\rangle$ excited state, the system adiabatically follows a dark state (see below) and acquires no phase. A subsequent $\pi$ pulse restores the control qubit state.

To show why the controlled phase evolution leads to the desired fan-out gate, assume for simplicity that all the targets are initially in state $|1_t\rangle$ and, hence, in the state $|i_c\rangle \otimes (\frac{|0_t\rangle -|1_t\rangle}{\sqrt{2}})^{\otimes k}=2^{-k/2}|i_c\rangle \otimes  \sum \limits_{j=0}^{k}  \left(\begin{array}{c} k\\ j \end{array}\right)(-1)^j |1_t\rangle^{\otimes j}|0_t\rangle^{\otimes k-j}$ after application of the first Hadamard gate. The phase acquired by each component in this state during the $2\pi$ pulse is $(-1)^j$ if the control atom is in state $|1_c\rangle$ and unity if the control is in state  $|r_c\rangle$. The $|1_c\rangle$ component becomes
$|1_c\rangle \otimes  2^{-k/2}\sum \limits_{j=0}^{k}  \left(\begin{array}{c} k\\ j \end{array}\right) (-1)^{2j} |1_t\rangle^{\otimes j}|0_t\rangle^{\otimes k-j}=|1_c\rangle \otimes (\frac{|0_t\rangle +|1_t\rangle}{\sqrt{2}})^{\otimes k}$, which after the second Hadamard target operation results in the C-NOT$^{k}$ gate.

\section{Dark states}
\label{sec_dark states}
The resonant excitation in presence of the exchange interaction between control and target atoms results in the formation of a multi-qubit dark state.
In the analyses in this section we disregard the variation of the dipole interaction strengths $B_1$ and $B_2$ due to the different distances between the atoms. This permits an effective treatment of the problem in a reduced basis of symmetric states.  The full model is assessed by numerical simulations with atoms on a regular lattice.

{\bf Toffoli gate :} Our gate relies on the dipole-dipole interaction $|r_c r_t\rangle {\stackrel{B_1}{\rightleftharpoons}} |a_cb_t\rangle$ which may be tuned into exact resonance by application of external fields. The always resonant intra-component exchange interaction $|r_c b_c\rangle {\stackrel{B_2}{\rightleftharpoons}} |b_cr_c\rangle$ will yield a perturbation on our dark state dynamics and must also be taken into account.

\begin{figure}
\centering
\raggedleft%\centering
\scalebox{0.45}{\includegraphics*[viewport=100 0 900 480]{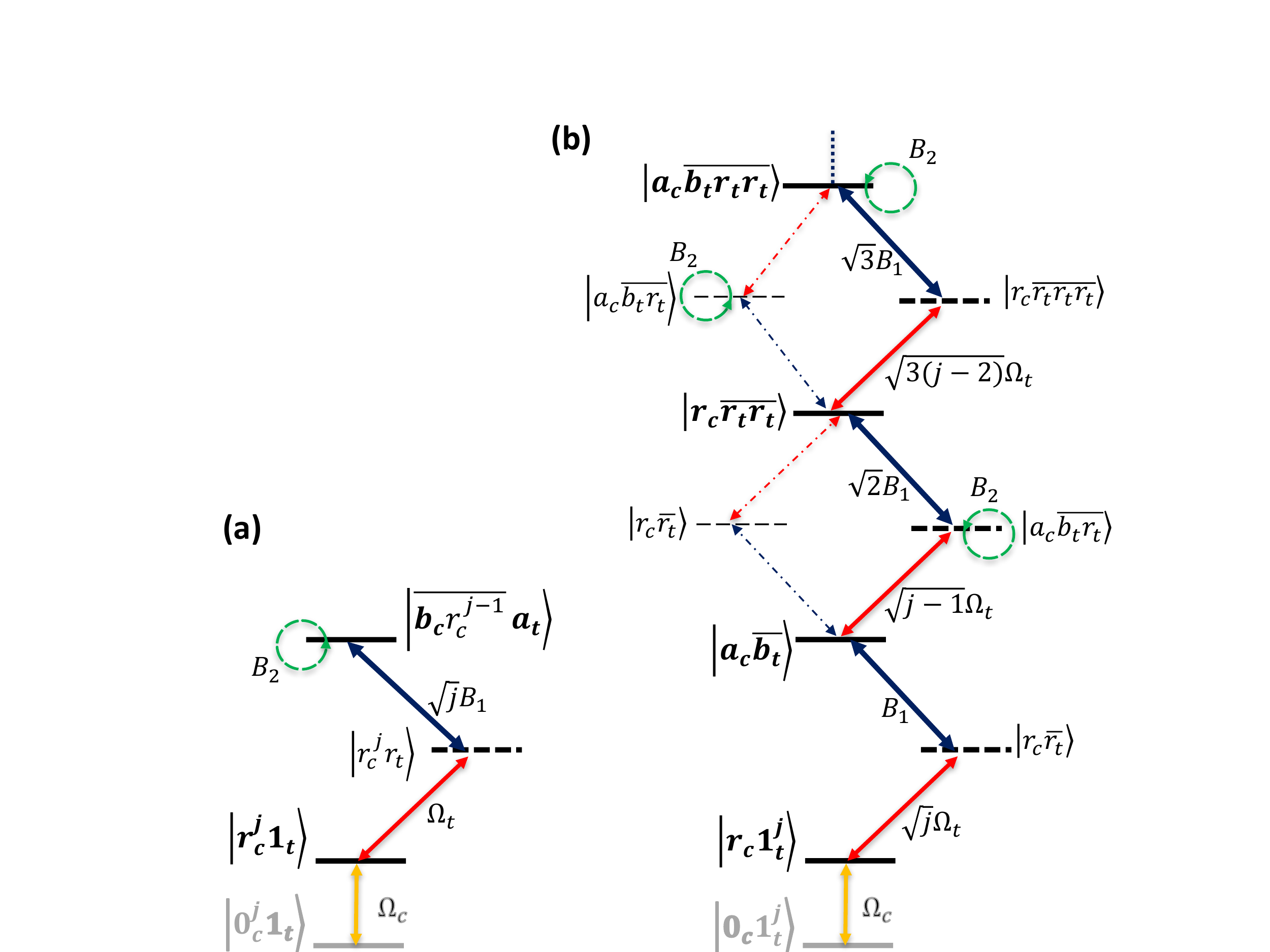}}
\caption{Dark states in (a) Toffoli and (b) fan-out gates. Only the state components shown in bold face get populated. In the fan-out gate (b) the system undergoes up to $j$ STIRAP transfer processes. The collective states' notation indicated by overline, is defined in the text.
}\label{FigDarkState}
\end{figure}

We thus split the Hamiltonian into the following two parts, ($\hbar=1$)
 \begin{eqnarray}
 &&V_{cc}=B_2\sum_{i<l} (|b_{c_i} r_{c_l}\rangle\langle r_{c_i} b_{c_l}|+\text{h.c.}) \\ \nonumber
&&H_{d}=\Omega_t/2(|r_t\rangle\langle 1_t|+\text{h.c.})+\sum_{i=1}^k B_1(|b_{c_i}a_t\rangle\langle r_{c_i} r_t|+\text{h.c.}), \\ \nonumber
 \end{eqnarray}
with atomic indices $i$ and $l$. Subject to $H_d$, during slow turn on of the target Rabi frequency $\Omega_t$, the system populating initially a state with $j$ Rydberg excited control atoms and no target excitations, $|r_c^{ j} 1_t\rangle$, acquires a component of the state $|\overline{b_c r_c^{ j-1}}a_t\rangle$, forming a superposition similar to the ``dark state'' in STIRAP processes \cite{Una98}, i.e., the coupling to the intermediate state $|r_c^{ j} r_t\rangle$ vanishes due to destructive interference of the transition amplitudes,
\begin{equation}
\label{ToffoliDark}
|d\rangle_t=\cos(\theta_t)|r_c^{ j} 1_t\rangle - \sin(\theta_t) |\overline{b_c r_c^{ j-1}}a_t\rangle,
\end{equation}
where $\tan(\theta_t)=\frac{\Omega_t}{2\sqrt{j}B_1}$
  and $|\overline{b_c r_c^{ j-1}}a_t\rangle$ is the symmetric state with one of the $j$ atoms transferred from the $r_c$ to the $b_c$ Rydberg level. The coherent coupling of states is illustrated in Fig.~\ref{FigDarkState}a.

While Eq.~\ref{ToffoliDark} is a zero energy eigenstate of $H_d$, the intra-component interactions $V_{cc}$  disturb the state and contribute a loss of fidelity together with other error mechanisms that we shall assess in detail in the following sections and in the appendix. We can estimate the error due to $V_{cc}$ to be of magnitude $\bar{P}_{ |\overline{b_c r_c^{j-1}}a_t\rangle }B_2\tau_t=  \frac{\pi \Omega_t B_2}{4jB_1^2+\Omega_t^2}$ for $j>1$ where $\bar{P}_{ |\overline{b_c r_c^{j-1}}a_t\rangle}$ is the average population of the interacting state over the duration $\tau_t$ of the $2\pi$ rotation of the target atom. As expected this error is negligible in the interesting regime of Fig.~\ref{FigTotalError}.

\begin{figure}
\centering
    \subfloat{%
    \includegraphics[width=.24\textwidth]{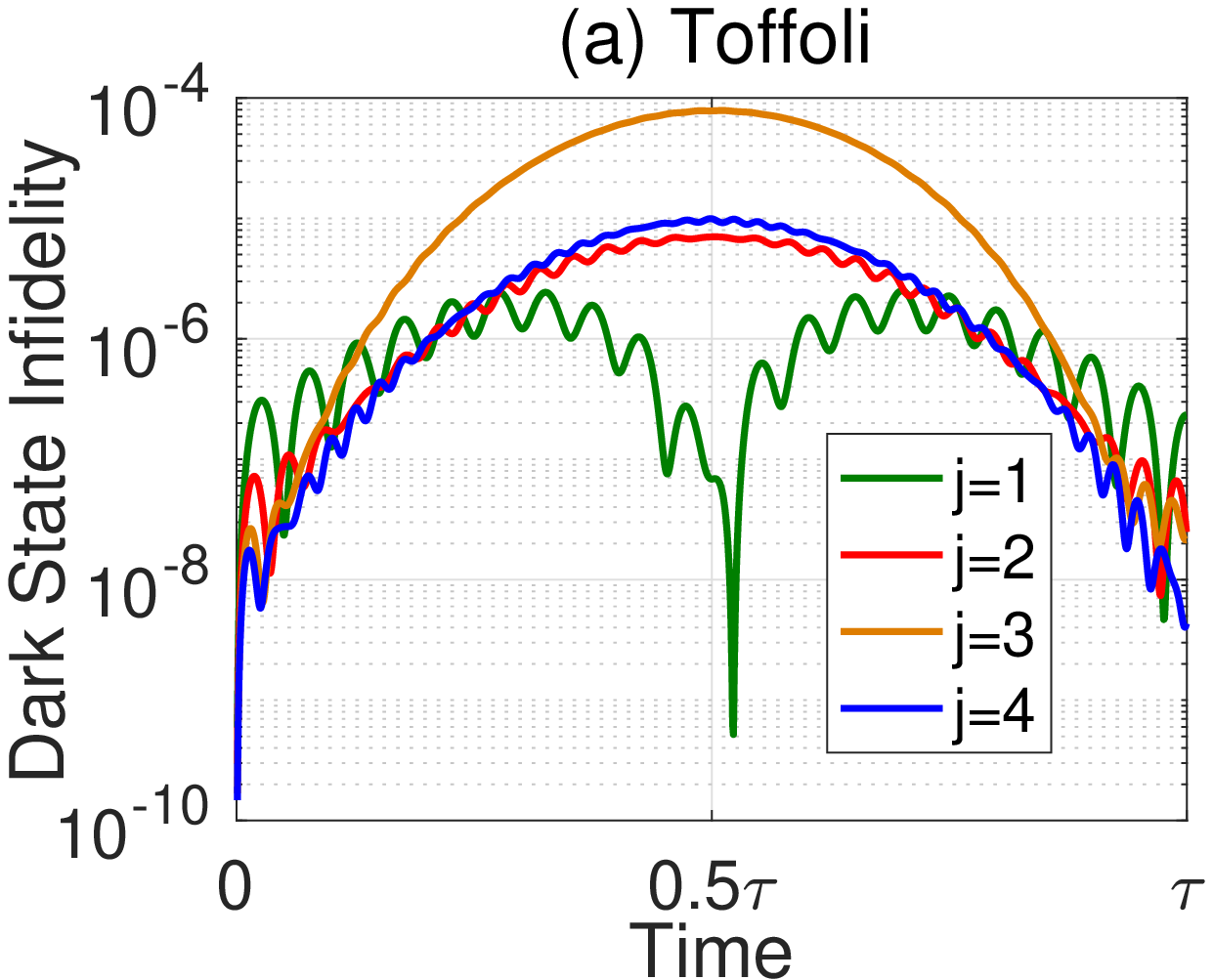}}\hfill
    \subfloat{%
\includegraphics[width=.24\textwidth]{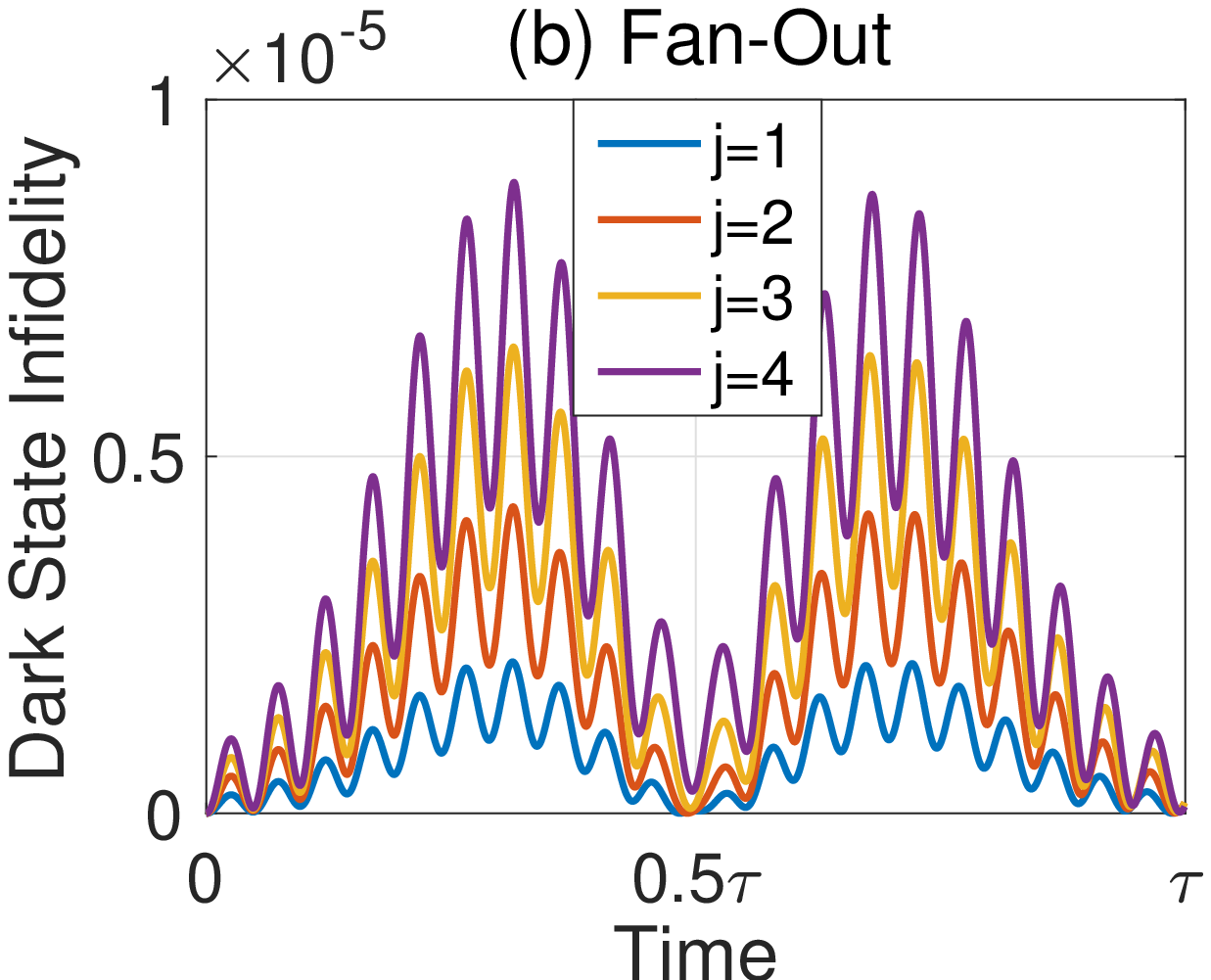}}\hfill
\caption{The figure shows the infidelity of the numerical solution to the time dependent Schr\"odinger equation  with respect to the time dependent dark eigenstate (2) for the Toffoli gate (a), and  (4) for the fan-out gate (b). The Rydberg qubit states are $|101S,109S\rangle$  with interaction strengths listed in table I in App. A1. The target Gaussian $2\pi$ pulse with duration ${\tau}$ and maximum Rabi frequency $\Omega_{t}=0.1B_1$ is applied to atoms on a square lattice with lattice constant of 10$\mu$m. The atomic system is chosen to maximize the values of $B_1$ and minimize the ones of $B_2$.
}\label{FigDarkStateEvol}
\end{figure}

We have tested the validity of the adiabatic eigenstate (Eq.~\ref{ToffoliDark}) of $H_d$ by solving the time dependent Schr\"odinger equation for the system subject to $H_d$ and $V_{cc}$ with realistic experimental parameters and with atoms occupying a regular lattice geometry. As shown in Fig.~\ref{FigDarkStateEvol}a, the infidelity is very small  and the initial state is retrieved with minimal non-adiabatic loss \cite{FigCap}.
Appendix 2  presents further discussion of the adiabatic following.

{\bf Fan-out gate:}
To explain the dark state that is formed from control-target interaction in the fan-out gate, we consider the case where initially a single control qubit is excited to $|r_c\rangle$ with a $\pi$ pulse. In the next step we apply a $2\pi$ pulse which is resonant with the
transition $|1_t\rangle \leftrightarrow |r_t\rangle$ of the target atoms, while the system is subject to the dipole-dipole interactions $|r_c r_t\rangle {\stackrel{B_1}{\rightleftharpoons}} |a_cb_t\rangle$ and $|r_t b_t\rangle {\stackrel{B_2}{\rightleftharpoons}} |b_tr_t\rangle$ with the couplings $B_1$ and $B_2$.
The Hamiltonian is thus
 \begin{eqnarray}
 &&H=\sum_{i=1}^k [\Omega_t/2(|r_t\rangle_i\langle 1_t|+\text{h.c.})+\\ \nonumber
&&B_1(|a_cb_t^i\rangle\langle r_c r_t^i|+\text{h.c.})]+B_2\sum_{i<l} (|b_t^i r_t^l\rangle\langle r_t^i b_t^l|+\text{h.c.}).
 \end{eqnarray}

Starting from the initial state with $j$ target atoms in $|1_t\rangle$, the state evolves adiabatically under the dark state interference mechanism shown in Fig.~\ref{FigDarkState}(b).
The figure shows a ladder of state components (bold solid lines), occupied  with amplitudes that ensure vanishing coupling to the rightmost ladder of states (dashed lines).
Here $|r_c {\overline{r_t^{m}}}\rangle$ is the normalized sum of all possible configurations with $m$ target atoms in the Rydberg state and $j-m$ target atoms remaining in the state $|1_t\rangle$, while $k-j$ atoms are in the uncoupled state $|0_t\rangle$. In the $m^{th}$ two-photon step, the coupling strengths are given by $[\sqrt{m(j-m+1)}\Omega_t/2,\sqrt{m}B_1]$ and $[\sqrt{(m-1)(j-m+1)}\Omega_t/2,\sqrt{m}B_1]$ for odd and even $m$, respectively, The corresponding fan-out dark state reads
\begin{eqnarray} \label{EqDarkState}
  &&|d\rangle_{f}=\sum \limits_{i=0}^{j/2}\cos(\theta^f_{(j-2i)!})\sin(\theta^f_{(2i)!})      	|r_c \overline{r_t^{2i}}\rangle   \\ \nonumber
  &&-\sum \limits_{i=0}^{j/2-1}\cos(\theta^f_{(j-2i-1)!})\sin(\theta^f_{(2i+1)!})	 |a_c \overline{b_t  r_t^{2i}}\rangle,
 \end{eqnarray}
where $\tan(\theta^f_{m=2i})=\frac{\Omega_t}{2B_1} \frac{\sqrt{j-m+1}\sqrt{m-1}}{\sqrt{m}}$ and $\tan(\theta^f_{m=2i+1})=\frac{\Omega_t}{2B_1} \frac{\sqrt{j-m+1}\sqrt{m}}{\sqrt{m}}$ and $\tan(\theta^f_{m!}) \equiv \prod \limits _{l=1}^m \tan(\theta^f_{l})$ with $\tan(\theta^f_{0!}) \equiv 1$. Fig.3(b) shows that the time dependent solution of the Schr\"odinger equation follows the adiabatic eigenstates very well.
The probability of exciting $m$ Rydberg atoms is given by $P_m=(\tan\theta^f_{m!})^2$.
Since the states are less populated for higher $m$,
the most important contribution of the target interaction comes from $|a_c {\overline{b_t r_t^{2}}}\rangle$ populated with probability $(\tan\theta^f_{3!})^2$.  The perturbation of the dark state by the exchange interaction is quantified by $\tfrac{j(j-1)(j-2)}{2} (\frac{\Omega_t}{2B_1})^6 B_2$, which has negligible effect in the regime of interest of Fig.~\ref{FigTotalError}.

\section{Error estimates}
\label{Sec_Error}
In this section we address the effects of spontaneous emission, errors in population rotations and non-adiabatic dynamics on the gate fidelity. In Appendices 4 and 5 we further discuss the effects of atomic motion and of non-resonant Rydberg channels leading to unwanted phases and population loss.
We first provide simple estimates of the errors and  in Fig.~\ref{FigTotalError} we compare with the more elaborate results obtained in Appendix A3 with the atoms positioned on a real lattice.

{\bf Toffoli:} The Rydberg state decay rate is given by $\Gamma$ \cite{Bet09}.
Over each $2\pi$ pulse between ground state and Rydberg level, the average decay and thus error probability of an atom is given by $\frac{2\pi}{\Omega}\Gamma$. Similarly, an atom has a probability to decay if it is maintained in an excited state during a gate.
In the Toffoli gate, the control atoms are not blocking each other and all   $j$  atoms  in state  $|0_c\rangle$ get excited to the Rydberg level. The maximum target Rydberg state population in the dark state is $\frac{\Omega_t^2}{4jB_1^2}$ when there are $j$ control atoms in the Rydberg level and $1$ when $j=0$. Therefore, the average spontaneous emission errors from target and control atoms are estimated by
\begin{eqnarray}
&&E_{se,t}= \frac{2\pi \Gamma}{\Omega_t}\frac{1}{2^{k+1}}[1+\sum \limits_{j=1}^k  \left(\begin{array}{c} k\\ j \end{array}\right) \frac{\Omega_t^2}{4jB_1^2}]\\ \nonumber
&&E_{se,c}=(\frac{2\pi}{\Omega_c}+\frac{4\pi}{\Omega_t})\Gamma\frac{1}{2^{k}}\sum \limits_{j=1}^k  \left(\begin{array}{c} k\\ j \end{array}\right) j =(\frac{2\pi}{\Omega_c}+\frac{4\pi}{\Omega_t})\frac{k\Gamma}{2}.
\end{eqnarray}
where $\Gamma \simeq 1$kHz at $T=77$K is the maximum decay rate of the applied Rydberg levels with $n\simeq 100$.
During population rotation between the ground and Rydberg levels, another error of magnitude $\frac{(j-1)^2D_{cc}^2}{\Omega_c^2}$ appears for each control atom due to the  unwanted interaction $D_{cc}=\frac{C^{mm}_6}{r_{cc}^6}$ between the control atoms where the interaction coefficient  $C^{mm}_6$ is provided for a set of states in Table I.
Finally, nearby Rydberg levels detuned by $\delta_{r=c,t}=[U(n)-U(n-1)]=Ry/n^3$ where $Ry$ is the Rydberg energy cause rotation errors, and we summarize the error contributions,
\begin{eqnarray}
&&E_{r1,c}=\frac{1}{2^{k}}\sum \limits_{j=2}^k  \left(\begin{array}{c} k\\ j \end{array}\right) j\frac{(j-1)^2D_{cc}^2}{\Omega_c^2}=\frac{k^3-k}{8}\frac{D_{cc}^2}{\Omega_c^2}\\ \nonumber
&&E_{r2,c}=\frac{1}{2^{k+1}}\sum \limits_{j=1}^k  \left(\begin{array}{c} k\\ j \end{array}\right) \frac{ \Omega_c^2}{4(\delta_c\pm (j-1) D_{cc})^2}\\ \nonumber
&&E_{r2,t}=\frac{1}{2} \frac{ \Omega_t^2}{4\delta_t^2}.
\end{eqnarray}
More detailed discussions of these estimates are presented in appendix A3.

Adiabatic manipulation of $\Omega_t$, prevents the scattering of population from the dark into the bright states and ensures return of the atomic population to the qubit basis. Non-adiabaticity during the Toffoli gate is quantified in  Appendix 2 by the error
\begin{equation}
E_{adi}=\frac{1}{2^{k+1}}\sum \limits_{j=1}^k  \left(\begin{array}{c} k\\ j \end{array}\right) \frac{\Omega_t^4}{640 B_1^4j^2}.
\label{Eq_AdiErrorTo}
\end{equation}
The trade off between errors $E_{se}$, $E_{r1}$ and $E_{adi}$, $E_{r2}$ yield optimal laser intensities that fulfill $D_{cc},\Gamma \ll \Omega_c \ll \delta_c$ and $\Gamma \ll \Omega_t \ll B_1,\delta_t$.

\begin{figure}
\centering
     \subfloat[]{%
    \includegraphics[width=.24\textwidth]{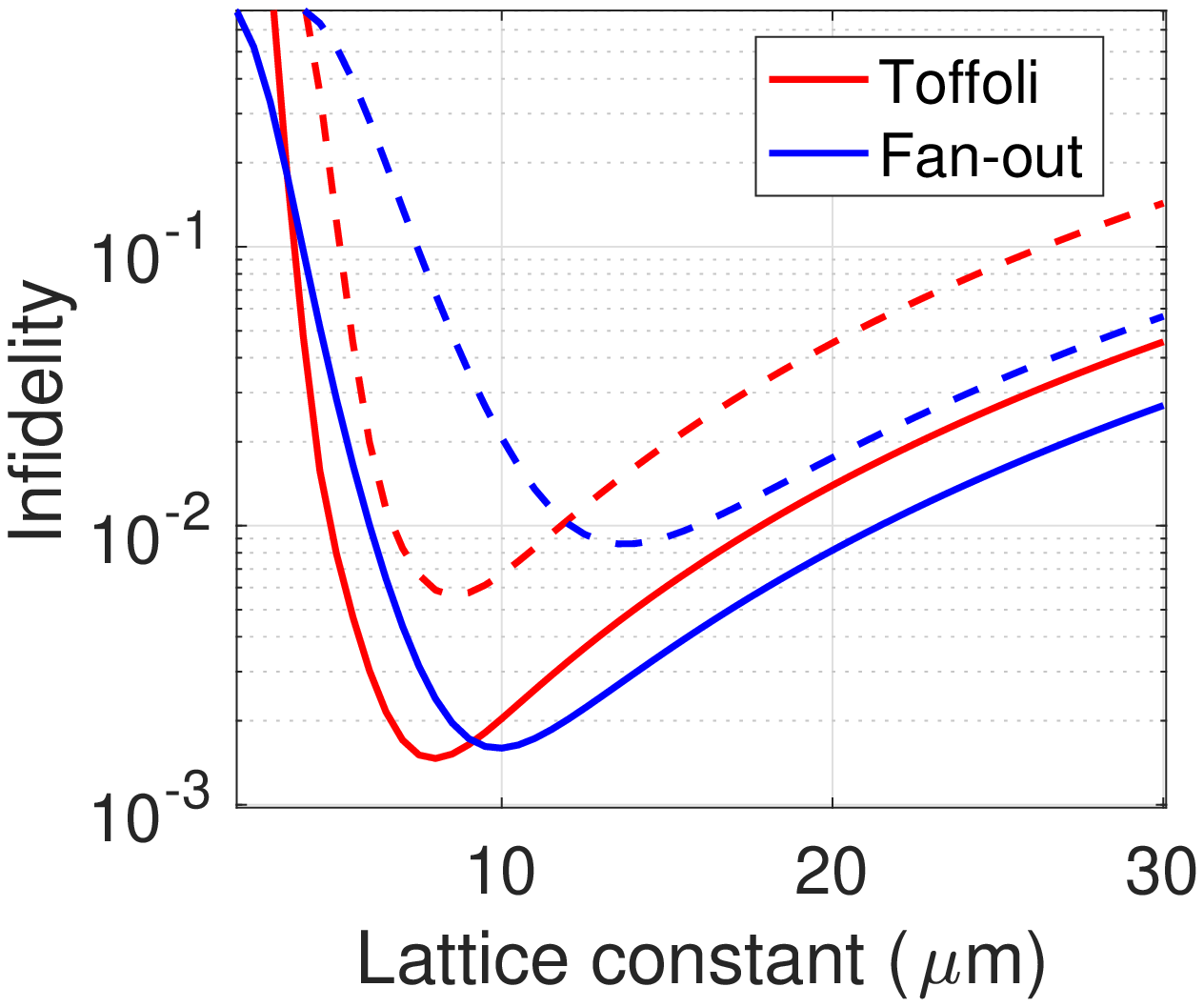}}\hfill
\subfloat[]{%
  \includegraphics[width=0.24\textwidth]{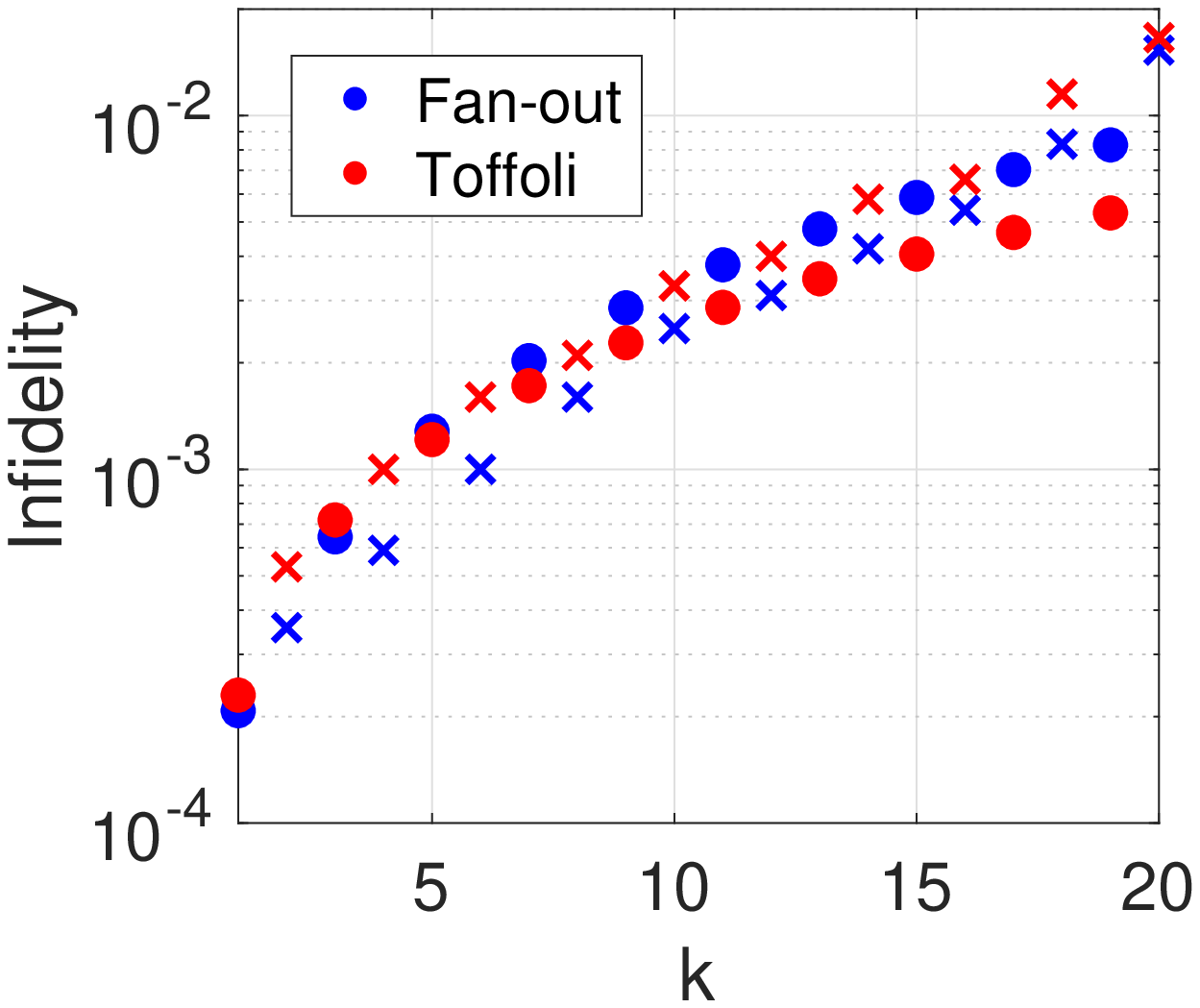}}\hfill
\caption{Total infidelity of dark state gates vs (a) atomic separation and (b) number of qubits $k$  (we use Rydberg states $|101S,109S\rangle$ of Cs atoms with intra and inter component interaction coefficients given in table I in App. A1).  In panel (a) gates with $k=6$  (solid lines) and $k=20$  (dashed lines) are carried out at the environment temperature of $T=77K$.  In panel (b) filled circles represent the analytical error estimate in Sec.~\ref{Sec_Error}, while crosses show more elaborate estimates determined with atoms located on a square lattice, see A3.  All the dynamical parameters are optimized for each point.
}\label{FigTotalError}
\end{figure}

{\bf Fan-out:}
The average errors in fan-out gates are quantified  along the same lines. Here the number of Rydberg excited target atoms is
$j$ for the case of $|1_c\rangle$, and in the case of $|0_c\rangle$ the population may be approximated by $P_t= \frac{j\Omega_t^2}{4B_1^2}$ in the weak driving regime.
\begin{eqnarray}
\label{Eq_ErrFan}
&&E_{se,c}=\frac{1}{2}(\frac{2\pi}{\Omega_c}+\frac{4\pi}{\Omega_t})\Gamma \\ \nonumber
&&E_{se,t}= \frac{2\pi \Gamma }{\Omega_t}\frac{1}{2^{k+1}}\sum \limits_{j=1}^k  \left(\begin{array}{c} k\\ j \end{array}\right)[j+ \frac{j\Omega_t^2}{4B_1^2}]\approx \frac{\pi k \Gamma }{2\Omega_t}\\ \nonumber
&&E_{r1}=\frac{1}{2^{k+1}}\sum \limits_{j=2}^k  j\frac{ [ (j-1) D_{tt} ]^2}{\Omega_t^2} \left(\begin{array}{c} k\\ j \end{array}\right)=\frac{k^3-k}{16}\frac{D_{tt}^2}{\Omega_t^2}\\ \nonumber
&&E_{adi}=\frac{1}{2^{k+1}} \sum \limits_{j=0}^k \left(\begin{array}{c} k\\ j \end{array}\right)\frac{j\Omega_{t}^4 } { 640 B_1^4}=\frac{k\Omega_{t}^4 } { 2560 B_1^4}\\ \nonumber
&&E_{r_2}=\frac{\Omega_c^2}{4\delta_c^2}+\frac{1}{2^{k+1}}\sum \limits_{j=1}^k  \left(\begin{array}{c} k\\ j \end{array}\right)[\frac{j\Omega_t^2}{4(\delta_t\pm (j-1)D_{tt})^2}+\frac{j \Omega_t^2}{4\delta_t^2}], \nonumber 
\end{eqnarray}
where in $E_{se,t}$ and $E_{r_2}$ the elements in bracket correspond to  $|1_c\rangle$ and $|0_c\rangle$ states. Adiabatic errors are derived in the appendix A2, and $D_{tt}=\frac{C^{mm}_6}{r_{tt}^6}$ is the intra-component interaction between target atoms at separation $r_{tt}$.

The total infidelity of the dark state gates vs number of qubits $k$ operating on a square lattice of Cs atoms are plotted in Fig.~\ref{FigTotalError}, where the circles represent the analytical error estimates \cite{Point2} and crosses report  the average lattice dependent error from a more detailed calculation with atoms positioned in a square lattice, see appendix 3.
 The targeted Rydberg states are $|101S,109S\rangle$, see  table I in  Appendix A1 for the corresponding interaction strengths. All dynamical parameters are optimized. The upper limit  $\Omega_t/B_1<0.42$ is considered to minimally perturb the energy of the dark  states and to fulfill the adiabatic error scaling, see A2.  In calculating $E_{se}$ the environment temperature of $T=77k$ is considered.  Room temperature performance is quantified in Fig.~\ref{FidVsn} of the appendix A1. A lattice constant of $r>8\mu$m is sufficient to preserve the dark state, avoid population leakage to non-resonant Rydberg pairs and keep the associated unwanted phases small, see Appendix A4.
The infidelities shown with cross symbols in Fig.~\ref{FigTotalError}b are obtained with the optimized dynamical parameters  $16$kHz $ <\Omega_t/2\pi <8$MHz, $\Omega_c/2\pi=16$MHz  and a lattice constant of $r=8\mu m$ for the Toffoli gate, and  $1$MHz $ <\Omega_t /2\pi<8$MHz,  $\Omega_c/2\pi=16$MHz and $8\mu m  <r<10\mu m$ for the Fan-out gate.
Following the parameter optimization, Fig.  \ref{FigTotalError} shows that the dark state multi-qubit gates can be realized for up to k=20 atoms with less than 1\% errors, making them suited for entanglement generation schemes and surface code error correction \cite{Wan11}.

The adiabatic dark state evolution improves the fidelity compared with the blockade scheme by reducing different rotation errors and by decoupling the motional degrees of freedom, see Appendix A5. The conventional blockade schemes entails a blockade leakage error $E_{r3}$, while the dark state approach  suppresses rotation errors in Eq.~\ref{Eq_AdiErrorTo} and \ref{Eq_ErrFan}  by a factor $\frac{E_{adi}}{E_{r3}}=\frac{\Omega_t^2}{160 B_1^2}$, see Appendix A2.
Also, in the blockade schemes the control-target interaction results in level shifts that enhance off resonant excitation of neighbouring Rydberg pairs with resulting gate errors   $E^{blo}_{r2,t}=\frac{1}{2^{k+1}} [\sum \limits_{j=1}^k  \left(\begin{array}{c} k\\ j \end{array}\right) \frac{ \Omega_t^2}{4(\delta\pm j B_{ct})^2}+ \frac{ \Omega_t^2}{4\delta^2}]$ in the Toffoli gate and $E^{blo}_{r_2,t}=\frac{1}{2^{k+1}}\sum \limits_{j=1}^k  \left(\begin{array}{c} k\\ j \end{array}\right)[\frac{j\Omega_t^2}{4(\delta\pm (j-1)D_{tt})^2}+\frac{j \Omega_t^2}{4(\delta \pm B_{ct})^2}]$ in the Fan-out gate. In comparison with the errors of the dark state gates, shown in Fig.4, the blockade Toffoli scheme working with simultaneous pulses results in the error range of $0.02<E<0.09$ for $3<k<24$  \cite{Ise11} and in the range  $0.015<E<0.35$ for the Fan-out gate for the same range of $k$ numbers.  Comparing with the filled circles in fig. 4b, the dark state multi-qubit gates reduce the infidelity by one to two orders of magnitude while operating at less demanding Rabi-frequencies.

\section{Implementation with superconducting circuits}\label{sec:sup}
\begin{figure}
\centering
\raggedleft
\scalebox{0.7}{\includegraphics*[viewport=0 170 400 440]{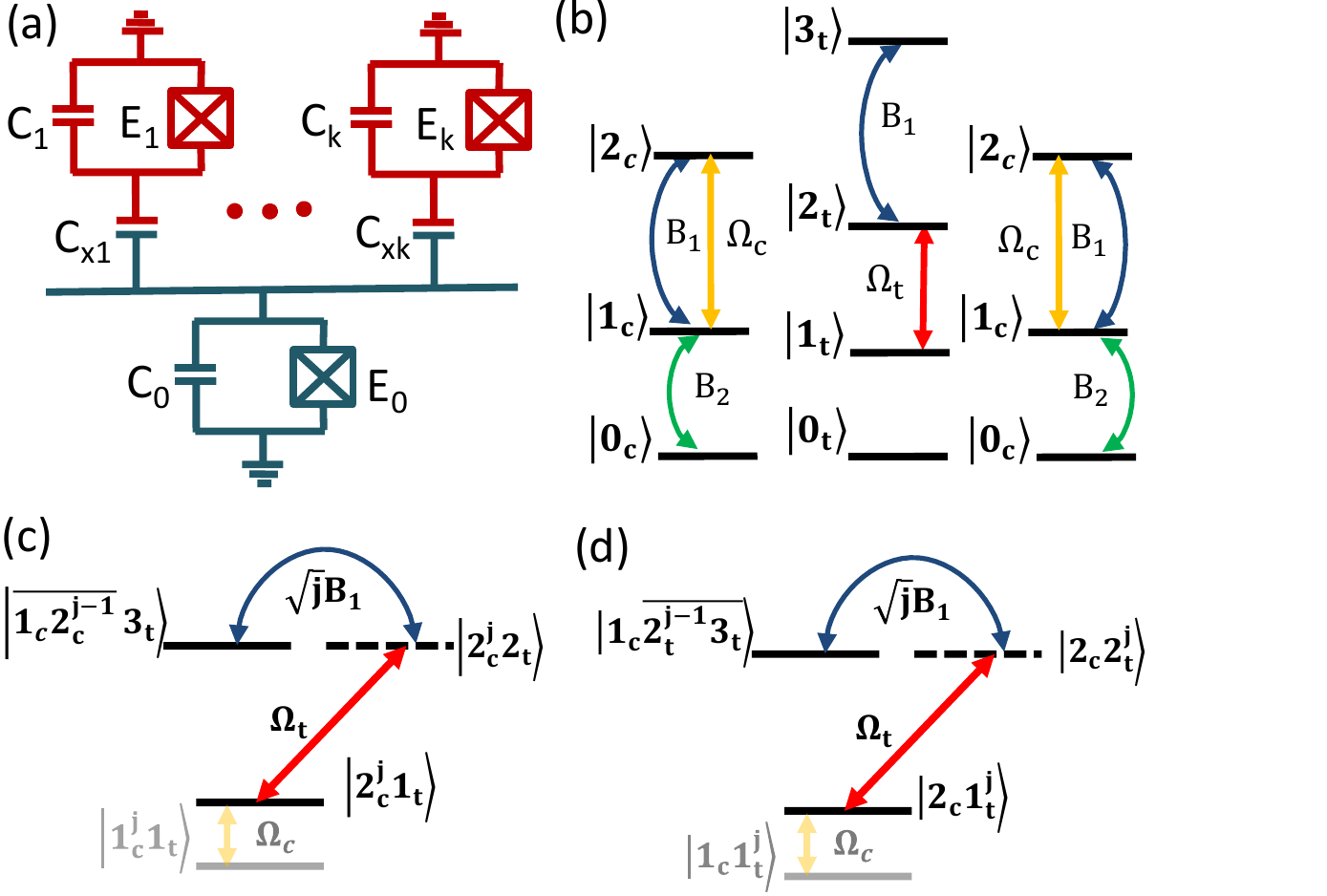}}
\caption{Implementation of adiabatic multi-qubit gates with superconducting circuits. (a) k control(target) qubits and a single target(control) qubit applied in the multi-qubit Toffoli(fan-out) gate. (b) Level structure of two control and one target qubit systems for the Toffoli gate, showing the resonant exchange processes with strengths $B_1$ and $B_2$. (c)((d)) Coupling of levels leading to multi-qubit dark superposition states (\ref{EqdarkSupCon}) in the Toffoli (fan-out) gates where only the state components shown in bold face are populated. The overline symbols in the state kets are explained in the text. }\label{FigSupCon}
\end{figure}

In this section we show that the adiabatic following of excitation-exchange eigenstates, analyzed in detail for Rydberg excited atoms, can also be employed for multi-qubit gates in superconducting circuit architectures \cite{Mot18}, see Fig.~\ref{FigSupCon}a.
%The excitation-exchange dark-state mechanism can also be realized with transmon qubits \cite{Mot18}. Here we design appropriate superconducting circuit architecture for the implementation of adiabatic multi-qubit gates, see Fig.~\ref{FigSupCon}a.
 The role of control(target) qubits in the Toffoli(fan-out) gates is held by the upper red circuit elements, $1 \ ... \ k$, and the target(control) qubit is shown as the lower blue circuit element. Fig.~\ref{FigSupCon}b shows the qubit level structure for the Toffoli gate, with logical qubit states $|0_{(c,t)}\rangle$ and $|1_{(c,t)}\rangle$ and auxiliary states $|2_{(c,t)}\rangle$ and $|3_t\rangle$.
The circuit parameters are chosen to make the product states $|3_t 1_c\rangle $ and  $|2_t2_c \rangle $  degenerate and coupled by strength $B_1$, while minimizing the resonant exchange coupling strengths $B_2$ between the control qubits.

The gate operation is similar to the atomic implementation discussed in Sec.~\ref{sec:Scheme}.  The Hamiltonian of the system while applying the classical drive in resonance with the $|1_t\rangle - |2_t\rangle$ transition is given by
\begin{eqnarray}
\label{HamSupCon}
&&H_t=\frac{\Omega_t}{2}(|1_t \rangle \langle 2_t | + \text{h.c.})+\sum \limits_{i=1}^k B_1 (|3_t 1_c^i\rangle \langle 2_t2_c^i | + \text{h.c.})\\ \nonumber
&&H_f=\sum \limits_{i=1}^k [ \frac{\Omega_t}{2}(|1_t^i \rangle \langle 2_t^i | + \text{h.c.})+B_1 (|3_t^i 1_c\rangle \langle 2_t^i 2_c | + \text{h.c.})],
\end{eqnarray}
for Toffoli and fan-out gates respectively (see details in App.6). 
In the presence of  at least one control and one target excitation, the evolution of the dark states during the target 2$\pi$ pulse is  given by
\begin{eqnarray}
\label{EqdarkSupCon}
&&|d\rangle_t=\cos(\theta)|2_c^{j} 1_t\rangle   -\sin(\theta) |\overline{1_c2_c^{j-1} }3_t \rangle \\ \nonumber
&&|d\rangle_f=\cos(\theta)|2_c 1_t^j\rangle  -\sin(\theta)  |1_c \overline{2_t^{j-1}3_t} \rangle
\end{eqnarray}
in the Toffoli and fan-out gates, respectively, see Fig.~\ref{FigSupCon}c,d. Like in the atomic implementation discussed above, the over-line symbols represent the normalized sum of states where one of the $j$ (control)target atoms is (de-)excited and  $\tan(\theta)=\frac{\Omega_t}{2\sqrt{j} B_1}$.
In Appendix  6, we briefly present how the qubit interaction parameters are obtained from the circuit capacitances and Josephson energies, and we discuss the multi-qubit gate fidelities achievable with realistic physical parameters.

\section{Conclusion}

In this article \cite{Khaz20} we have proposed and analyzed multi-qubit gates based on adiabatic evolution of Rydberg excited multi-atom dark states, formed by an interference between coherent driving terms and resonant dipole-dipole excitation transfer among pairs of atoms. Previous works \cite{Pet17} have shown that such dark state dynamics holds the potential to achieve better error scaling for two-qubit gates than the conventional Rydberg blockade mechanism, and our work has demonstrated the viability of the same mechanism for many atoms.
We show that simple estimates of the errors give rise to acceptable gate fidelities, and that one may conceivably apply the gate to up to 20 atoms to prepare multi-qubit entangled states by very short and fast laser pulse sequences.
Similar performance is predicted in this paper for an implementation with superconducting qubits. Schemes using a similar mechanism may be employed for Toffoli and fan-out multi-qubit gates on trapped ions (To be published).

For quantum computing and error correction, the fan-out and Toffoli gates are useful, and already for just 2-4 target and control qubits, multi-qubit gates that employ adiabatic following of exchange interaction eigenstates may have advantages over sequential operation of one and two-qubit gates, \cite{Ise11,Mol11,Pet16}. Such gates are both much faster and include fewer pulses and may hence have higher fidelity \cite{Gul15} (a C$_{20}$-NOT gate based on concatenated one- and two-qubit Rydberg gates would require about 690 sequential laser pulses addressing individual sites \cite{She09,Mas03}).

Toffoli and fan-out gates play key roles in quantum error correction \cite{Nie11,Cor98}, the Grover search algorithm \cite{Mol11} and Shor's factoring algorithm \cite{Van01}, and their implementation by few operations will impact the prospects for  fault tolerant quantum computing. The multitude of theoretical proposals for quantum computing and quantum simulations making use of higher order interactions \cite{Gla14,Cel19}  provide promising targets for the gates presented in this article on both superconducting architectures and atoms in regular spatial configurations in 1D \cite{Ber17,Omr19}, 2D \cite{Zha11,Pio13,Nog14,Xia15,Zei16,Lie18,Nor18,Coo18,Hol19,Sas19} and 3D \cite{Wan16,Bar18},

\section*{Acknowledgment }
The authors acknowledge financial support from Iran National Elites Foundation (MK), the Villum Foundation and the U.S. ARL-CDQI program through cooperative Agreement No. W911NF-15-2-0061 (KM).
Also (MK) wishes to thank the IPM-HPC centre for technical support.

\section*{Appendix}

\section*{A1: Interactions between Rydberg excited atoms}
\label{SecRealization}

Here we identify candidate Rydberg levels that are useful for the proposed gate implementation.  We denote by subscript $m$ the state associated with the $k\geq 1$ control (target) qubits in the C$_k$-NOT (C-NOT$^k$) gate. The state occupied by the single target (control) qubit is represented here by subscript $s$.
A natural choice of Rydberg states would be $r_{m}=a_s=nS_{1/2}$ and $r_{s}=b_m=nP_{3/2}$ with a resonant control-target exchange interaction.

However, in addition to unwanted coupling to other pairs of Rydberg states \cite{Pet17}, with this choice, the control and target qubits may  exchange their excited state components and hence ruin the multi-qubit gate performance.
To preserve the strong inter-component interaction, we use  Stark shifted resonant Rydberg pairs instead of excitation exchange resonant Rydberg pairs.
\begin{figure}
\centering
    \subfloat{%
    \includegraphics[width=.23\textwidth]{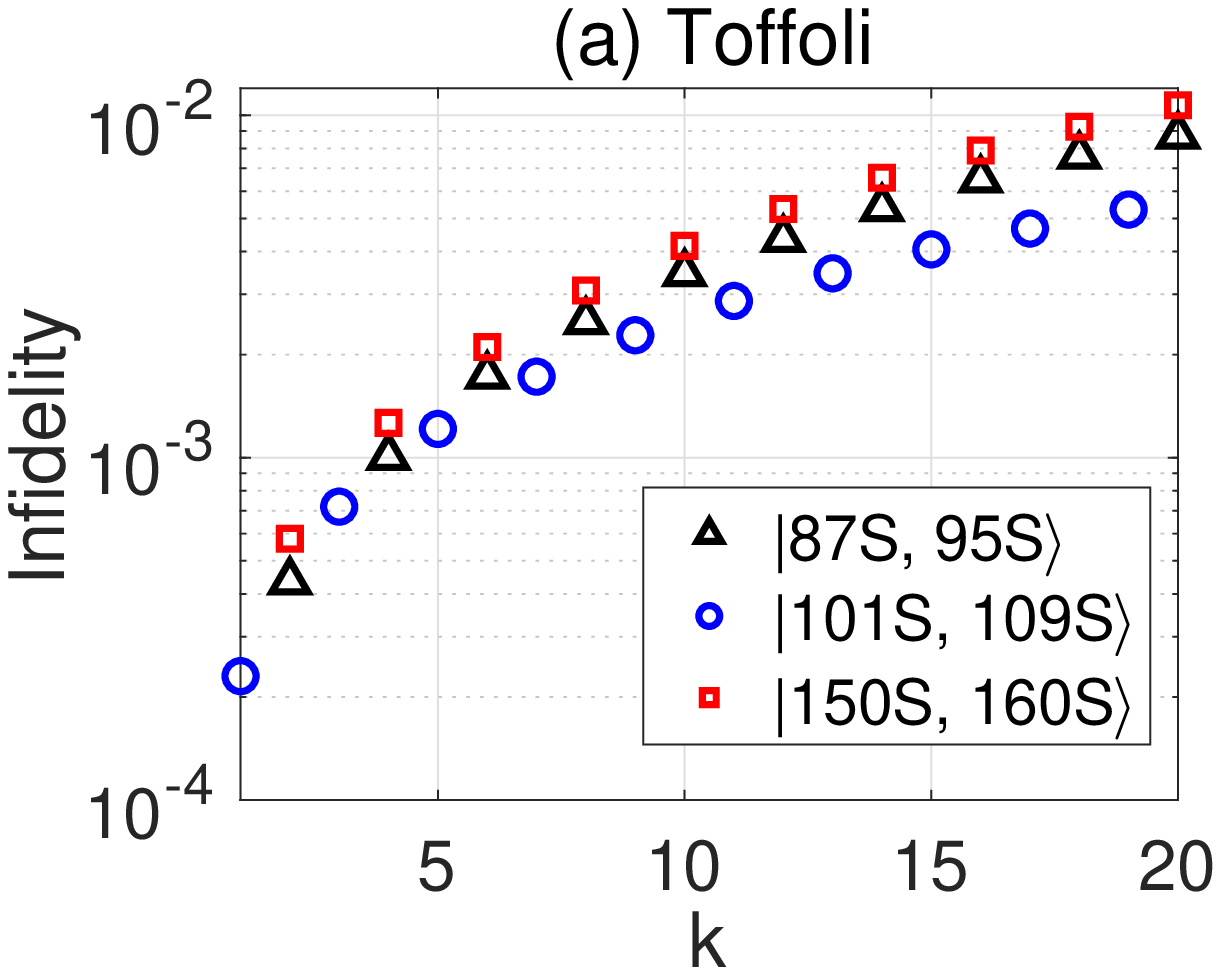}}\hfill
    \subfloat{%
    \includegraphics[width=.23\textwidth]{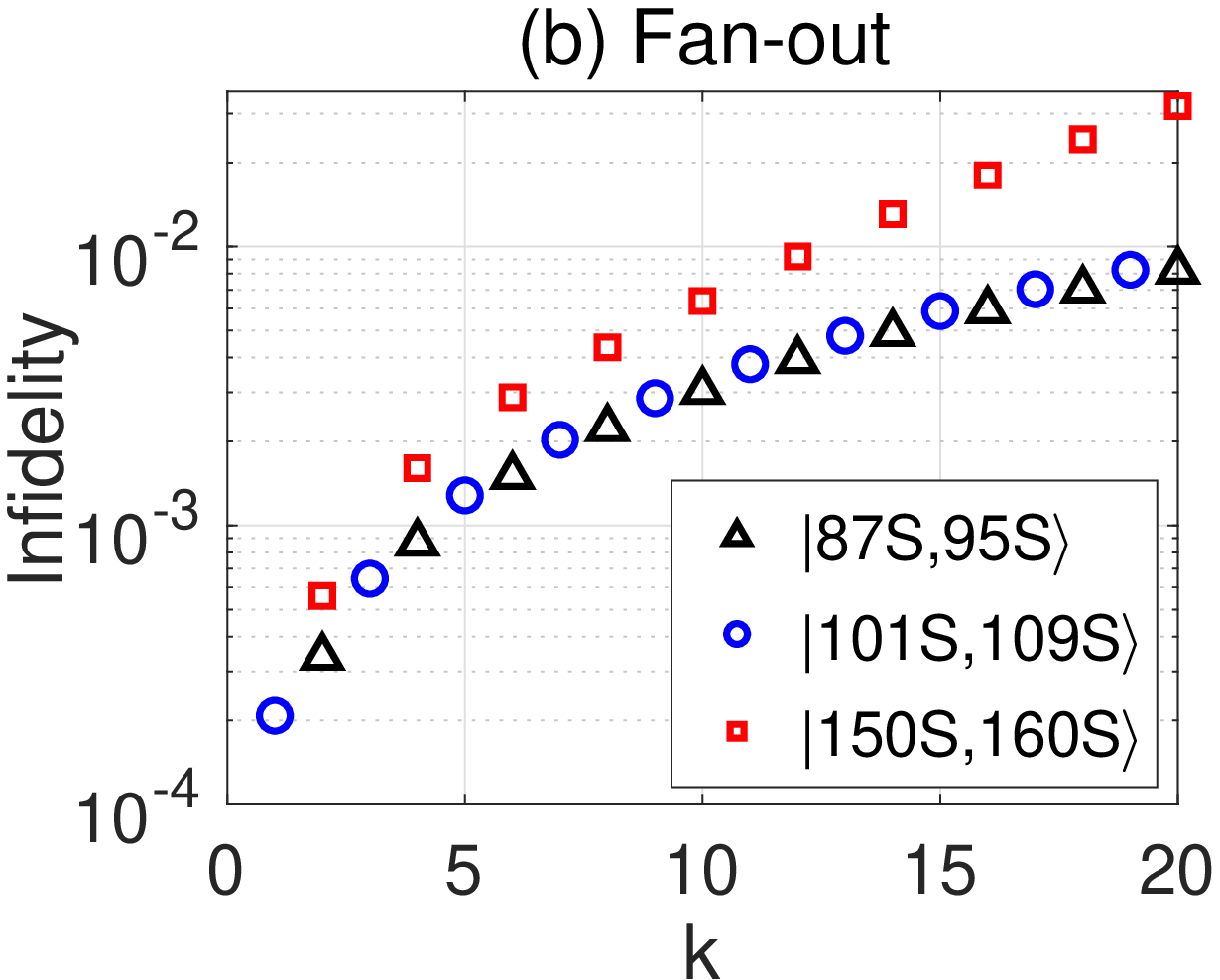}}\hfill
    \subfloat{%
    \includegraphics[width=.23\textwidth]{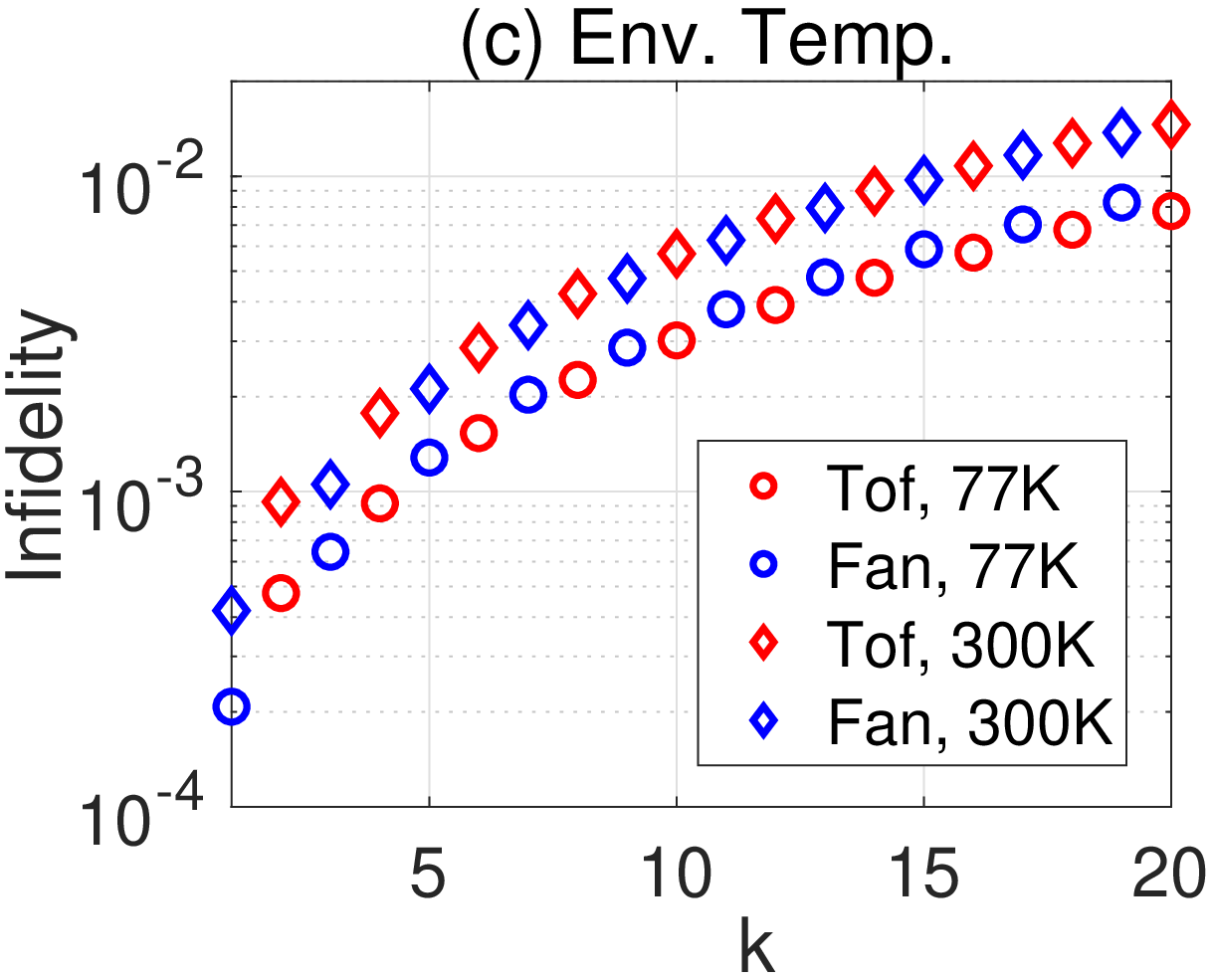}}\hfill
\caption{Fidelity of dark state gates as a function of qubit number $k$  for (a,b) different Rydberg states presented in table I and for (c) different environment temperatures. The environment temperature in (a,b) is $T=77$K and the Rydberg states in (c) are $|101S,109S\rangle$. The range of dynamic parameters are stated in the text.
}\label{FidVsn}
\end{figure}
Therefore, we shall choose different states and use an electric field to tune only the desired states into resonance. Applying the external electric field perpendicular to the planar array of atoms preserves isotropic interaction among atoms and it also improves the ratio of the inter-component interaction $C_3^{B_1}$  to the unwanted intra-component interaction $C_6^{mm}$ between multi-qubit states, see Sec.~\ref{Sec_Error}.

Rydberg atom pairs $|r_{m}r_{s}\rangle$ and $|a_{m}b_{s}\rangle$ are chosen to be in-resonance in the presence of the external field $E$.  Interaction coefficients of states $|r_{s}r_{m}\rangle$, $|r_{m}b_{m}\rangle$ and $|r_{m}r_{m}\rangle$ are represented by $C_{3}^{B_1} $, $C_{3}^{B_2} $ and $C_{6}^{mm} $ respectively.
The Stark shifts of the levels and the Van der Waals interaction coefficients $C_{6}^{mm} $ are calculated with perturbation theory for a range of Rydberg pairs within $\pm 3$ variation in principal quantum numbers and angular momentum $0<l<4$. Table I represents three candidate level schemes with different principal numbers  in Cs. Similar level schemes could be applied in Rb with the application of stronger electric fields.
When choosing the levels we have optimized the  $C_3^{B_1}/C_{6}^{mm}$ to reduce rotation gate errors.  
\begin{widetext}
\begin{center}
\begin{tabular}{ c  c c c c c c cc}
\hline
 \hline
  $|r_{s}\rangle$ & $|r_{m}\rangle$  & $|a_{s}\rangle$  & $|b_{m}\rangle$ &	$C_{3}^{B_1}$ & $C_{3}^{B_2}$ & $C_{6}^{mm} $    &$E $	 \\
     & &   &  &	${\scriptstyle(2\pi \text{GHz.}\mu m^3)}$ & ${\scriptstyle(2\pi \text{GHz.}\mu m^3)}$ & $ {\scriptstyle(2\pi \text{GHz.}\mu m^6)}$    &$ {\scriptstyle(\text{V/m})}$	 \\
 \hline
$95S_{1/2}1/2$\qquad \quad &  $87S_{1/2}1/2$\qquad \quad	 &      $95P_{3/2}3/2$\qquad \quad  &   87$P_{3/2}$3/2\quad   & -5.6 & -1.56 & -5  & 31.8  \\
  \hline
$109S_{1/2}1/2$\qquad \quad &  $101S_{1/2}1/2$\qquad \quad	 &      $109P_{3/2}3/2$\qquad \quad  &   $101P_{3/2}3/2$\quad   & -10.2 & -2.87 & -27.9  & 15.2  \\
   \hline
 $160S_{1/2}1/2$\qquad \quad &  $150S_{1/2}1/2$\qquad \quad	 &      $160P_{3/2}3/2$\qquad \quad  &   150$P_{3/2}$3/2\quad   & -49 & -14.3 & -4300  & 2  \\
\hline
\label{TableLattice}
\end{tabular}
\captionof{table}{Two Rydberg atom pairs $|r_{s}r_{m}\rangle$ and $|a_{s}b_{m}\rangle$ are in resonance in the presence of a static external field $E$. The multi-qubit state is represented by subscript $m$, associated with $k$ control (target) qubits in the C$_k$-NOT (C-NOT$^k$) gate. The single-qubit state is represented by subscript $s$. Interaction coefficients of states $|r_{s}r_{m}\rangle$, $|r_{m}b_{m}\rangle$ and $|r_{m}r_{m}\rangle$ are represented by $C_{3}^{B_1} $, $C_{3}^{B_2} $ and $C_{6}^{mm} $ respectively.  }
\end{center}
\end{widetext}
The fidelity of the dark state gates are compared for the three qubit states represented in table I and for different environment temperatures in Fig.~\ref{FidVsn}.  While $|101S,109S\rangle$ seems to be an optimum choice of state, moderate changes of the principal number are indeed possible \cite{Point3}.
The dynamic parameters for the states $|150S,160S \rangle$ are $d \in  [19\mu$m, $26 \mu$m$]$, $\Omega_c/2\pi \in [8$MHz, $9.5$MHz$]$, $\Omega_t/2\pi \in [1$MHz, $3$MHz$]$ for the Toffoli gate and $d \in  [20\mu$m$, 30 \mu$m$]$, $\Omega_c/2\pi$=10 MHz, $\Omega_t/2\pi \in [0.8$MHz, $2.4$MHz$]$ for the fan-out gate.
Choosing $|87S, 95S \rangle$ the range of parameters are
$d \in  [5\mu$m$, 7.5\mu$m$]$, $\Omega_c/2\pi=24$MHz, $\Omega_t/2\pi \in [5.5$MHz, $19$MHz$]$ for the Toffoli gate and $d \in  [5\mu$m$, 9.5\mu$m$]$, $\Omega_c/2\pi=24$MHz, $\Omega_t/2\pi \in [2.7$MHz, $19$MHz$]$ for the fan-out gate. Note that the critical distance defined in A4 is $d_c=4.5\mu m$ for this choice of states.
The range of parameters for T=300K in Fig.~\ref{FidVsn} are $d \in  [8\mu$m, $9.5 \mu$m$]$, $\Omega_c/2\pi $=24MHz, $\Omega_t/2\pi \in [5$MHz, 8MHz$]$ for the Toffoli gate and $d \in  [8\mu$m$, 12.5 \mu$m$]$, $\Omega_c/2\pi$=10 MHz, $\Omega_t/2\pi \in [2$MHz, $16$MHz$]$ for the fan-out gate.

 \section*{A2: Non-adiabatic errors}
  \label{SecAdiabaticity}
\begin{figure}
\centering
    \subfloat{%
    \includegraphics[width=.24\textwidth]{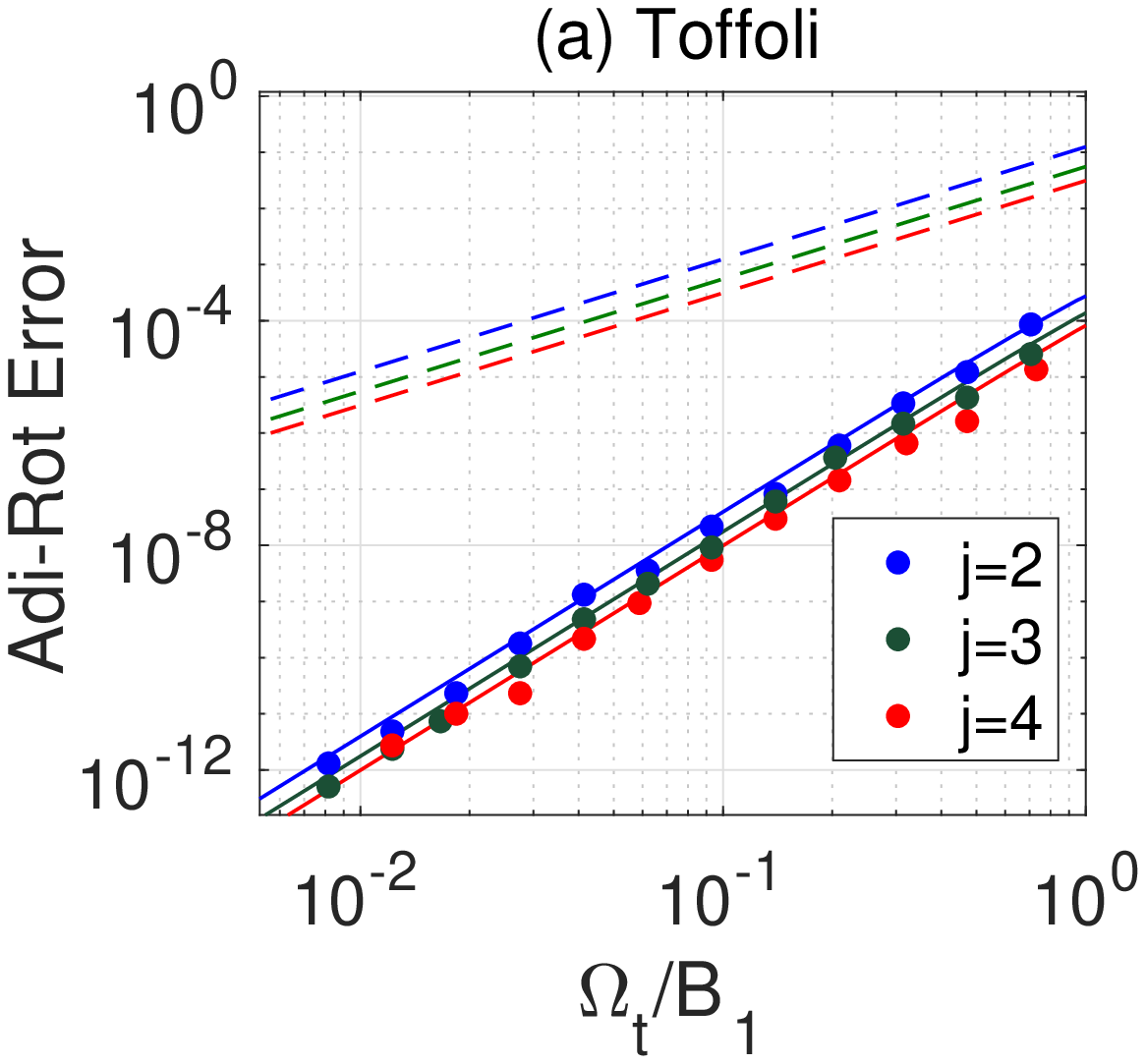}}\hfill
        \subfloat{%
    \includegraphics[width=.24\textwidth]{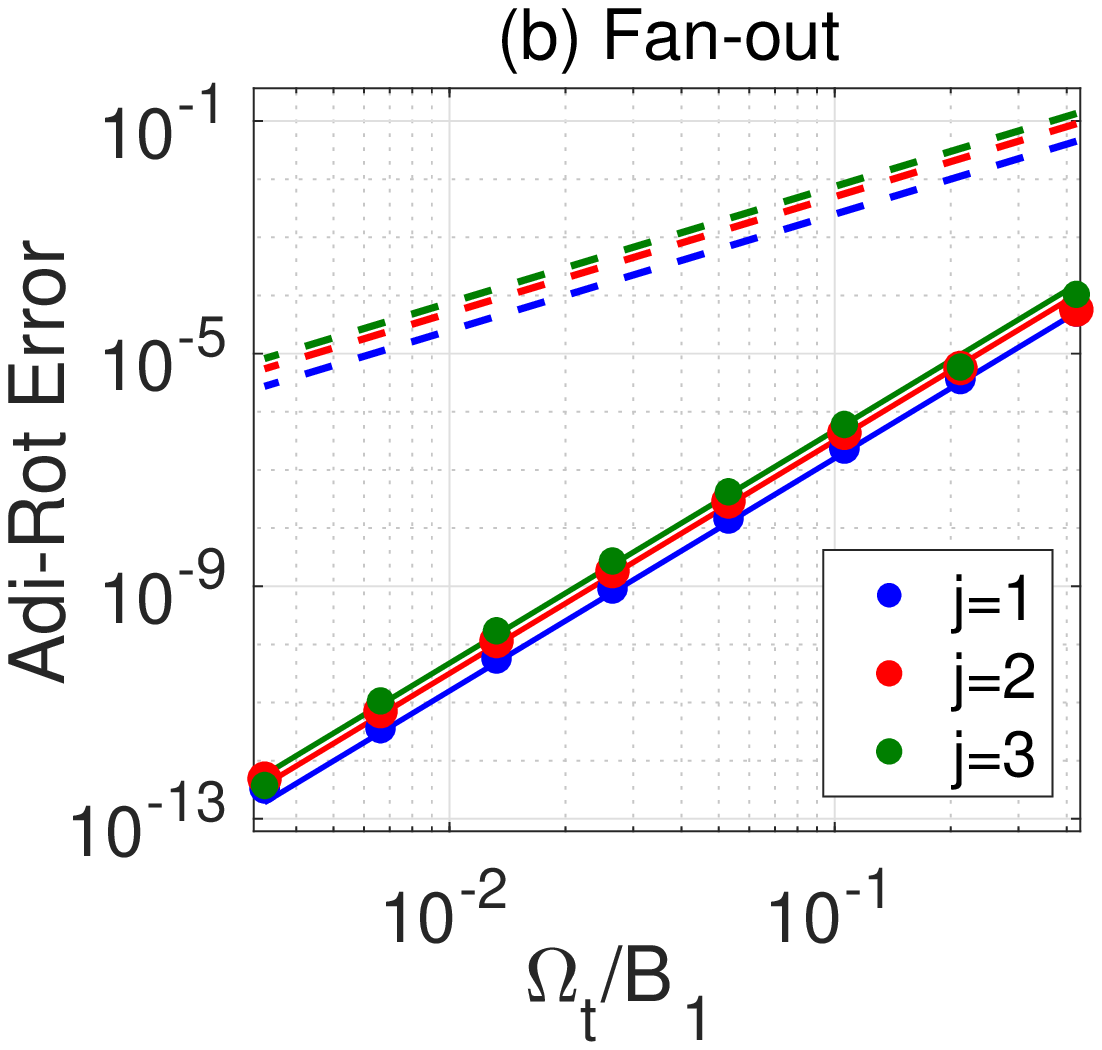}}\hfill
\caption{Non-adiabatic loss of population from the dark state in the (a) Toffoli and (b) Fan-out gates as a function of $\Omega_t/B_1$.  Scattered symbols represents results of numerical simulations with the atoms positioned on a square atomic lattice with a lattice constant of 10$\mu$m. We employ $|101S,109S\rangle$ Rydberg levels and we assume physical parameters maximizing (minimizing) the avrage value of $B_1$ ($B_2^{ave}$). The solid lines show the simple estimate of the  loss of population to the bright states, cf. Eqs.(\ref{EqAdiToffoli}, \ref{EqAdi}), multiplied by a factor of three.
The dashed lines show the rotation error of the corresponding blockade gates.
}\label{Adiabaticity}
\end{figure}

In this appendix we quantify the non-adiabatic errors in the Toffoli and Fan-out gates. Results obtained by numerical solution of the Schr\"odinger equation are shown with the filled symbols and compared with analytical estimates in Fig.~\ref{Adiabaticity}.

{\bf Analytical estimates for Toffoli gate:} The non-adiabatic loss of population from the dark state $|d\rangle_{t}$ in Eq.~\ref{ToffoliDark} during the Toffoli gate is estimated by $E_{adi}=\frac{\dot{\theta}_t^2}{( \Omega_t^2/4+j B_1^2)}$  where   $\dot{\theta}_t=\frac{\frac{\dot{\Omega}_t}{2\sqrt{j}B_1}}{1+\tan^2(\theta_t)}$.
Here we consider a Gaussian target pulse of  $\Omega(t)=\Omega_t(e^{-\frac{(t-T/2)^2}{2\sigma^2}}-e^{-\frac{(T/2)^2}{2\sigma^2}})$ with RMS width of $\sigma=T/5$ and the pulse duration of $T$ given by $\int \limits_0^T\Omega(t)dt=2\pi$.
The scattered population out of the dark state at the end of the target pulse is evaluated to
 \begin{equation}
 \label{EqAdiToffoli}
E_{adi}^{t}\approx  \frac{\Omega_{t}^4}{640\pi j^2 B_1^4}.
\end{equation}
In the conventional Toffoli blockade gate scheme, any blockade leakage population would directly affect the conditional phase and thus lead to an error,   $E^t_{r3}=\frac{\Omega_t^2}{4 j^2 B_1^2}$.
This value is plotted as the dashed lines in Fig.~\ref{Adiabaticity}a.

{\bf Analytical estimates for Fan-out gate:}
 In the weak driving regime $(\Omega_t/2 \ll B_1)$, the main population remains in the first three levels of the STIRAP process, see  Fig.~\ref{FigDarkState}b.
Restricting the Hamiltonian  to these levels, $H_{(m=1)}=\tfrac{\sqrt{j}}{2}\Omega(t) (|r_c1_t^j\rangle \langle r_c\bar{r}_t|+\text{h.c.})+B_1 (|a_c\bar{b}_t\rangle \langle r_c\bar{r}_t|+\text{h.c.})$, the dynamics in the time dependent eigenbasis of $H_{(m=1)}$ is governed by
\begin{equation}
\tilde{H}_{(m=1)}=\left(  {\begin{array}{c c c}
   \omega^+(t) & -i\frac{\dot{\theta}_1^f}{2} & 0 \\
   i\frac{\dot{\theta}_1^f}{2} & 0 & i\frac{\dot{\theta}_1^f}{2} \\
     0 & -i\frac{\dot{\theta}_1^f}{2} & \omega^-(t)
  \end{array} } \right)
 \end{equation}
where $\omega^{\pm}(t)$  and 0 are the eigenvalues of $H_{(m=1)}$, and $\tan(\theta_{(m=1)}^f)=\frac{\sqrt{j}\Omega(t)}{2B_1}$ is defined after Eq. 4.
To preserve the population in the dark eigenstate, the off-diagonal elements must be sufficiently smaller than the bright states energies $|\dot{\theta}_1^f(t)| \ll \sqrt{\tfrac{j}{4}\Omega^2(t)+B_1^2}$.
With the Gaussian target pulse with $\sigma=T/5$, we obtain the non-adiabatic error
 \begin{equation}
 \label{EqAdi}
E_{adi}^{f}\approx \frac{j\Omega_{t}^4}{640\pi B_1^4},
 \end{equation}
 at the end of the target pulse.
The corresponding rotation error in the higher two-photon steps in the EIT ladder of Fig.  \ref{FigDarkState}b  is multiplied by their excitation probability $(\tan \theta_{m!}^f)^2$ and do not contribute in the weak driving regime.

The numerical evaluation in Fig.~\ref{Adiabaticity} shows that the maximum loss of population to the bright states is governed by the non-adiabatic estimate in Eq.~\ref{EqAdiToffoli}, \ref{EqAdi} (the simple estimate has been corrected by a factor 3, which is compatible with the magnitude of the oscillations of the non-adiabatic population in Fig. 3.).
It is interesting to note that this error is again significantly smaller than the blockade gate rotation error  $E^f_{r3}=\frac{j\Omega_t^2}{4B_1^2}$,
 plotted with the dashed lines in Fig.~\ref{Adiabaticity}b.
The ratio of the errors in the adiabatic and blockade gates is given by $E_{adi}/E_{r3}\propto \frac{\Omega_t^2}{160 B_1^2}$, which suggests that the dark state approach may work at stronger driving regimes and hence allow faster operation.

%==============

\section*{A3: Gate errors for atomic configurations on a lattice}

Unlike in the error estimates in the main text, here we take into account that different qubit configurations with equal Rydberg atom numbers do not lead to same errors, due to the different interaction strengths over the lattice. We thus evaluate these interaction strengths for definite spatial configurations and we calculate the average gate fidelity over the $2^{k+1}$ qubit states. The main influence of the varying interactions occur in the rotation errors.

\begin{figure}
\centering
     \subfloat{%
    \includegraphics[width=.24\textwidth]{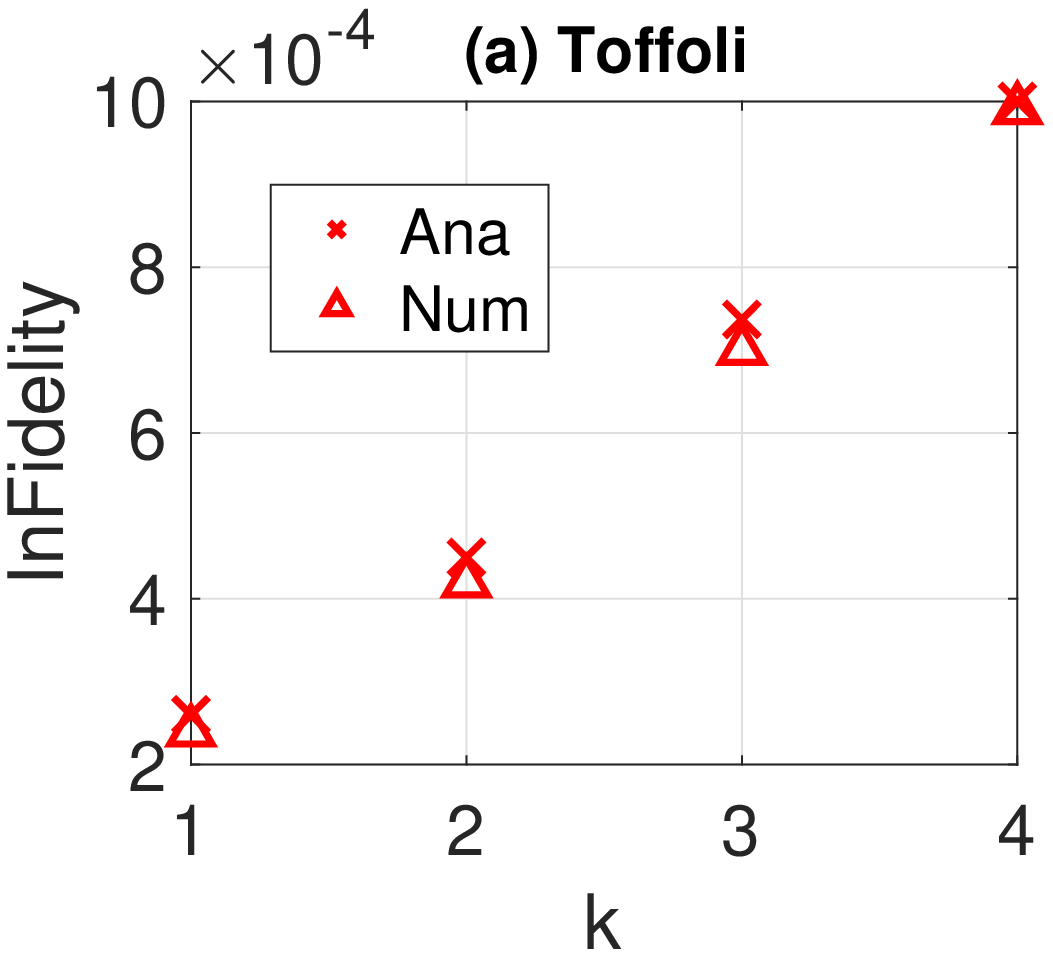}}\hfill
         \subfloat{%
    \includegraphics[width=.24\textwidth]{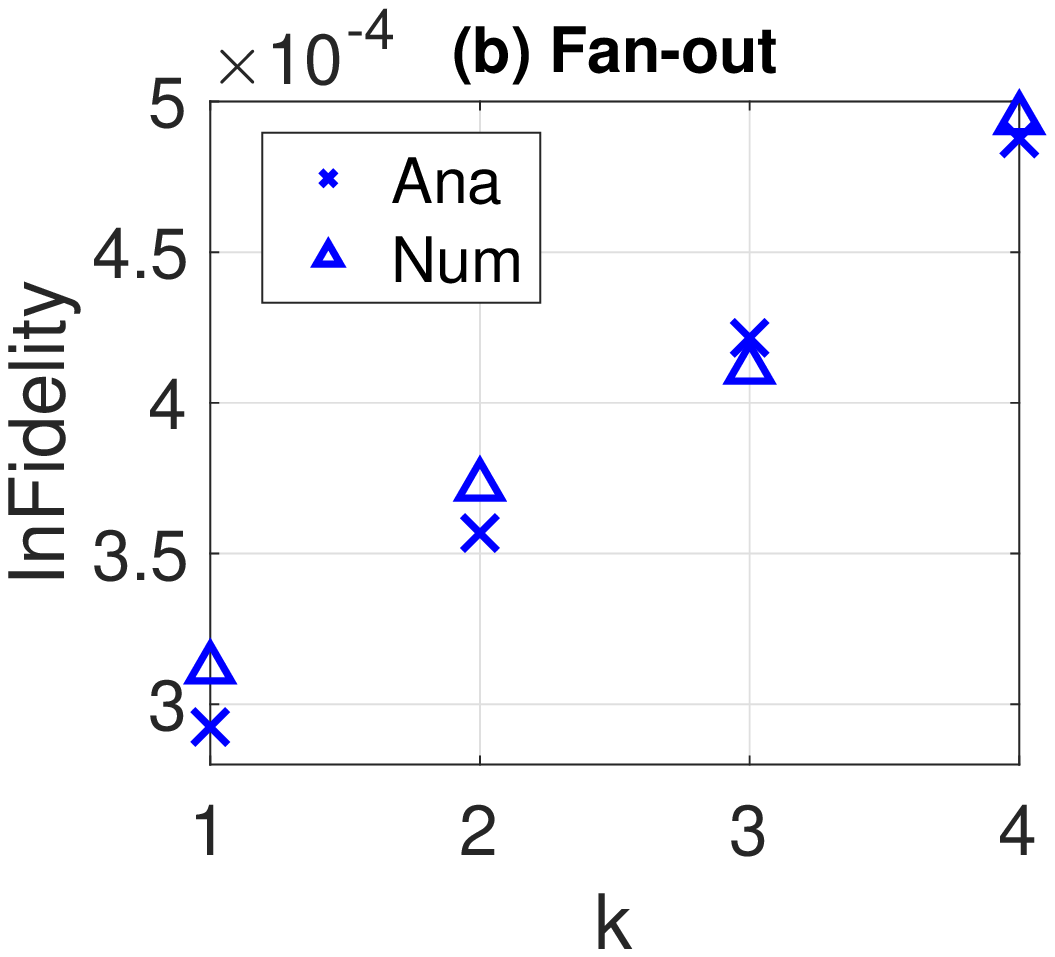}}\hfill
\caption{Comparison of analytical and numerical error calculations of multi-qubit gates over a square optical lattice.  In our  numerical simulations, the Schr\"odinger equation is solved on the tensor product space of the atoms.  The gates are operating on atoms populating neighboring positions in a square lattice with lattice constant of 8$\mu$m.  We assume an environment temperature of T=77K and laser coupling strengths $\Omega_t/B_1=0.42$ and $\Omega_c/2\pi=16$MHz, exciting the Rydberg states $|101S,109S\rangle$, see table I for the values of interaction parameters.
   }\label{Num}
\end{figure}

{\bf Numerical evaluation:}
For $k\le4$ it is possible to solve the Schr\"odinger equation numerically and quantify the gate fidelity averaged over all input qubit states \cite{Mol},
\begin{equation}
\label{FidDefenition}
F=[\text{Tr}(MM^{\dagger})+|\text{Tr}(M)|^2]/[n(n+1)]
\end{equation}
with $M=U_{id}^{\dagger}U_{gate}$, where $U_{id}$ and $U_{gate}$, represents ideal and realistic gate operations.  $U_{gate}$ is obtained from numerical simulation of the gate for all the possible $2^{k+1}$ qubit product state configurations, taking into account the atomic interactions imposed by the lattice geometry. In practice, we solve the Schr\"odinger equation on the tensor product space of (k+1) 5-level atoms (representing qubit and Rydberg levels, including extra Rydberg levels to simulate the second type of rotation errors estimated by  E$_{r_2}$ in the text). The results are depicted with the triangle symbols in Fig.~\ref{Num} for the parameters listed in the figure caption. The cross symbols are obtained by averaging analytical estimates for each classical qubit configuration, as described in the following.

{\bf Toffoli gate:}
In the $q$th qubit configuration ($1<q<2^{k+1}$), with $j_q$ control atoms occupying the $|0_c\rangle$ state, the first type of rotation  error
 is estimated by the sum of each control atom's error $E_{r1}^{(q)}=\sum\limits_{l=1}^{j_q}(\frac{\Delta_l}{\Omega_t})^2$  where $\Delta_l=\sum\limits_{m\neq l}^{j_q} D_{cc}(r_{lm})$ is the interaction of $l$th atom with all the $j_q-1$ other control atoms in the $|r_c\rangle$ state.
 The second type of rotation error from control atoms is estimated by $E_{r2,c}^{(q)}=\sum\limits_{l=1}^{j_q} \frac{\Omega_c^2}{4(\delta_c \pm \Delta_l)^2}$, where $\delta_c$ is the level spacing of the closest  accessible Rydberg level to $|r_c\rangle$, while an error contribution from the target atom is given by $E_{r2,t}^{(q)}=\frac{\Omega_t^2}{4\delta_t^2}$ when the target atom is in state $|1_t\rangle$ and zero otherwise.
Finally the average lattice dependent error is given by averaging over all qubit configurations.

{\bf Fan-out:}
In the $q$th qubit configuration, with $j_q$ target atoms that are in the $|1_t\rangle$ state, the  first type of rotation error for the $|1_c\rangle$ state
 is estimated by $E_{r1}^{(q)}=\sum\limits_{l=1}^{j_q}(\frac{\Delta_l}{\Omega_t})^2$ where $\Delta_l=\sum\limits_{m\neq l}^{j_q} D_{tt}(r_{lm})$ is the level shift of $l$th atom due to the $j_q-1$ other target atoms in the $|1_t\rangle$ state.
The second rotation error is estimated by $E_{r2}^{(q)}=\sum\limits_{l=1}^{j_q} (\frac{\Omega_t^2}{4(\delta_r \pm \Delta_l)^2}+\frac{\Omega_t^2}{4\delta_t^2})$, where $\delta_t$ is the level spacing of the closest Rydberg state to $|r_t\rangle$. The two terms in brackets correspond to $|1_c\rangle$, $|0_c\rangle$ states respectively. For the initial $|0_c\rangle$ state, the probability of exciting 2 target atoms is low in the regime of interest of Fig.~\ref{FigTotalError}, and hence target-target interactions do not contribute to errors. The main control-target interaction channel due to $B_1$ then populates the $|a_cb_t\rangle$ pairs as desired and does not cause rotation error.

The good quantitative agreement between the results of the full quantum evolution and the refined analytical estimates give confidence in the latter approach and qualifies its use for large atom numbers in Fig.4 of the manuscript.

 \section*{A4: Leakage to non-resonant Rydberg pairs}
 \label{smallPhases}
 \begin{figure}
\centering
     \subfloat[]{%
    \includegraphics[width=.24\textwidth]{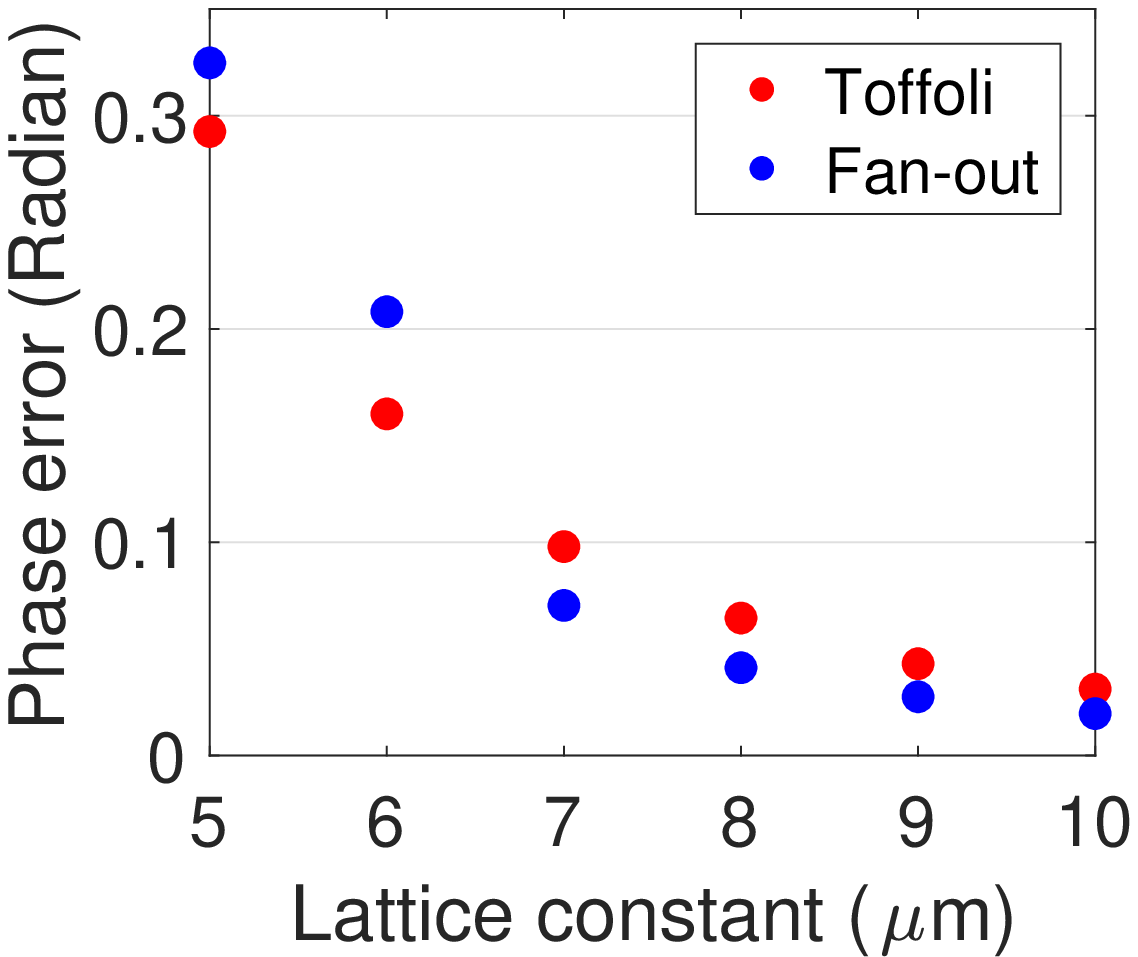}}\hfill
         \subfloat[]{%
    \includegraphics[width=.24\textwidth]{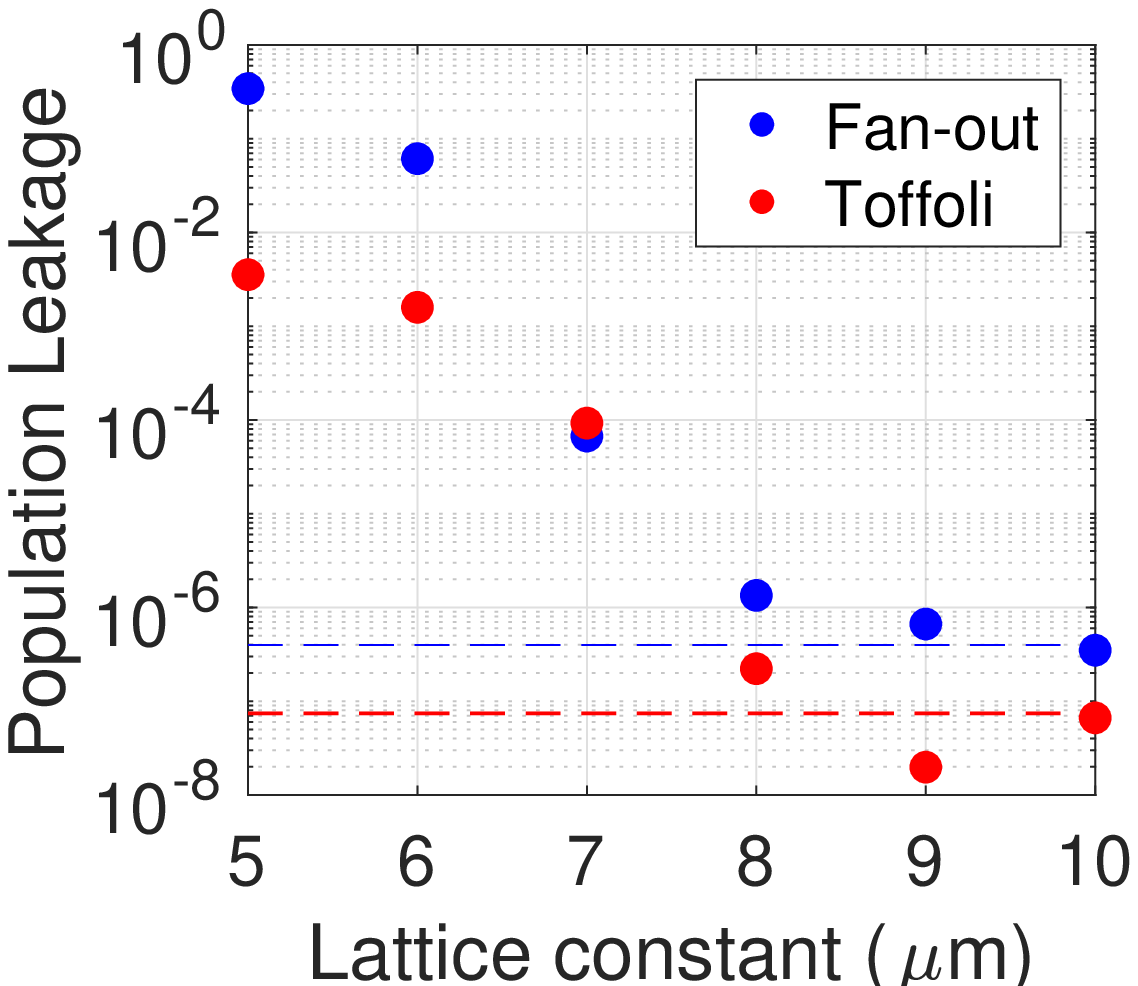}}\hfill
\caption{  Effects of non-degenerate channels (see table II) on (a) the acquired unwanted phase  and  (b) population leakage from the dark state after C-NOT$^2$  and C$_2$-NOT gates operation. The atoms assume a linear configuration with the single control or target qubit in the middle.
For distances larger than 8$\mu$m, the phase is negligible and the population leakage from the dark state is comparable to the non-adiabatic loss (dashed line). We assume laser coupling strengths $\Omega_t/B_1=0.1$ and  $\Omega_c/2\pi=16$MHz, exciting the Rydberg states $|101S,109S\rangle$
}\label{leakage}
\end{figure}

In our derivation of the dark states followed by our quantum system, we only included the main resonant exchange interactions $B_1$ and $B_2$ and neglected the weaker off-resonant channels.
The coupling to other non-resonant Rydberg pairs may lead to population loss and deviation from the ideal phase.
Here we recall the Rydberg states introduced in table I, $r_s=109S_{1/2}1/2$ $r_m=101S_{1/2}1/2$ $a_s=109P_{3/2}3/2$ $b_m=101P_{3/2}3/2$. With the atomic plane perpendicular to the quantization axis and with a $15.2\ V/m$ electric field along the quantization axis to make $|r_sr_m\rangle$ resonant with $|a_sb_m\rangle$, we obtain the physical couplings listed in table II.

To calculate the effects of the near-resonant Rydberg pairs in Fig.~\ref{leakage}, we fix $\Omega_t/B_1=0.1$ and simulate the Toffoli (C$_2$-NOT) and fan-out gate (C-NOT$^2$) gate operations with a 5 level single qubit atom and two control or target atoms with 9 levels (see table II).

The main reduction of fidelity comes from the third state listed in Table II.  Below a critical distance $d_c=\sqrt[3]{\frac{C_3}{\delta}}=8\mu m$, the coupling to that state is larger than its energy detuning and disturbs the dark state at the heart of the scheme.
Fig.~\ref{leakage}b shows that leakage out of the desired state becomes smaller than the adiabatic loss $E^{Fan}_{adi}=\frac{j\Omega_t^4}{640 B_1^4}$ and $E^{Tof}_{adi}=\frac{\Omega_t^4}{640 j^2B_1^4}$ (dashed lines) for lattice constants above 8$\mu$m. At lattice constants above 8$\mu$m, almost all the leakage channels experience weaker interactions and hence accumulate smaller unwanted phases. This is why we assume lattice constants larger than 8$\mu$m in the optimization of gate fidelities in Fig.~\ref{FigTotalError} in the main text.

\begin{flushleft}
\setlength{\extrarowheight}{3pt}
\begin{tabular}{|c c c c| }
\hline
No. &Coupled pairs  & $C_3/2\pi$ & $\delta/2\pi$ \\
 & & {\small GHz.$\mu$m$^3$} &  {\small MHz} \\
\hline
\hline
1&$ |r_sr_m\rangle \leftrightarrow |a_sb_m\rangle$ & -10.2 & 0  \\
2&$|r_mb_m\rangle \leftrightarrow |b_mr_m\rangle$ & -2.9 & 0\\
3&$ {|r_sr_m\rangle \leftrightarrow |109P_{1/2}\tfrac{-1}{2}, \, 101P_{3/2}\tfrac{-1}{2}\rangle}$ & 5 & 9.5  \\
4&$|r_mr_m\rangle \leftrightarrow |101P_{3/2}3/2, 101P_{3/2}3/2\rangle$ & -8.6 & 382  \\
5&$|a_sb_m\rangle \leftrightarrow |108D_{5/2}5/2,99D_{5/2}5/2\rangle$ & -6.5 & 52\\
6&$|a_sr_m\rangle \leftrightarrow |108D_{5/2}5/2,100P_{3/2}3/2\rangle$ & -14&-207 \\
7&$|r_mb_m\rangle \leftrightarrow |100P_{1/2}\tfrac{-1}{2}, 100D_{5/2}1/2\rangle$ & 3 & 3\\
\hline  \end{tabular}
\captionof{table}{List of near-resonant Rydberg pairs Subscripts $s$ and $m$ represents the single (e.g. target in Toffoli) and multiple atoms (e.g. controls in Toffoli).}
\end{flushleft}

\section*{A5. Motional degrees of freedom and gate errors}
The main dipole-dipole control-target interaction causes the system to follow a time dependent dark state with zero value and hence gradient of the energy with respect to the atomic spatial coordinates. As a result, despite the strong interactions, there will be no mechanical force between the atoms excited to Rydberg level. This is an  additional advantage of the current proposal as it eliminates (to leading order) any unwanted entanglement of the qubit states with the atomic motion.
Only non-adiabtaic corrections and the weak intra- and inter-component  interaction channels contribute a small correction to the ideal adiabatic evolution and hence the motional entanglement is suppressed.

\section*{A.6 Implementation with superconducting circuit}

The Lagrangian of the circuit in Fig.~\ref{FigSupCon}a is given by
\begin{equation}
L= \sum \limits_{i=0}^k [\frac{C_i}{2} \dot{\phi}_i^2+E_i \cos({\phi}_i)]+\sum \limits_{i=1}^k \frac{C_{xi}}{2}  (\dot{\phi}_i-\dot{\phi}_0)^2,
\end{equation}
where $\phi_i$ represent node flux variables.
The node charges  are the conjugate momenta of the node flux variables, $q_i=\frac{\partial L}{\partial \dot{\phi}_i}$.  Expanding the cosine function to fourth order yields the Hamiltonian
\begin{equation}
\hat{H}=\frac{1}{2} \vec{\hat{q}}^T{\bf C}^{-1} \vec{\hat{q}} - \sum \limits_{i=0}^k E_i(\frac{\hat{\phi}_i^2}{2}-\frac{\hat{\phi}_i^4}{24})
\end{equation}
where the capacitance matrix ${\bf C}$ has the form (for $k=2$)
 \begin{equation}
{\bf C}=\left(  {\begin{array}{c c c}
  C_0+C_{x1}+C_{x2} & -C_{x1} & -C_{x2} \\
  -C_{x1} & C_1+C_{x1} & 0 \\
     -C_{x2} & 0 &C_2+C_{x2}
  \end{array} } \right).
\end{equation}
We introduce oscillator raising and lowering operators through $\hat{\phi}_i=\sqrt{\frac{Z_i}{2}}(\hat{b}_i^{\dagger}+\hat{b}_i)$ and $\hat{q}_i=\frac{i}{\sqrt{2Z_i}}(\hat{b}_i^{\dagger}-\hat{b}_i)$ with the impedances, $Z_i=\sqrt{\frac{\mathbb{C}_{ii}}{E_i}}$, where the inverse capacitance matrix elements are defined as $\mathbb{C}_{ij}=({\bf C}^{-1})_{(i,j)}$.
The Hamiltonian can thus be written in terms of the bosonic raising and lowering operators
\begin{eqnarray}
\hat{H}=&& \sum \limits_{i=0}^k [(\frac{\mathbb{C}_{ii}}{2Z_i}+\frac{E_iZ_i}{2})\hat{b}_i^{\dagger}\hat{b}_i + \frac{E_iZ_i^2}{16} \hat{b}_i^{\dagger} \hat{b}_i^{\dagger} \hat{b}_i \hat{b}_i] \\ \nonumber
&&- \frac{1}{2} \sum \limits_{i>j=0}^k  \frac{\mathbb{C}_{ij}}{\sqrt{Z_iZ_j}} (\hat{b}_i^{\dagger}\hat{b}_j + \hat{b}_j^{\dagger}\hat{b}_i).
\end{eqnarray}
The first line defines  anharmonic ladders of energy levels while the second line represents excitation exchange. 
The energy of the n$^{th}$ level (i.e. n$^{th}$ Fock state) of the i$^{th}$ artificial atom is given by
\begin{eqnarray}
\omega_{n_i}=n_i\sqrt{E_i \mathbb{C}_{ii}} +\frac{\mathbb{C}_{ii}}{16} n_i(n_i-1).
\end{eqnarray}
We define the energy detuning between  product states of two artificial atoms $i$ and $j$ by $\delta_{n1_i,n1_j}^{n2_in2_j}=(\omega_{n2_i}+\omega_{n2_j})-(\omega_{n1_i}+\omega_{n1_j})$, and the anharmonicity of energy levels in unit $i$ is  $\alpha_{i}=(\omega_{2_i}-\omega_{1_i})-(\omega_{1_i}-\omega_{0_i})$.

Table III, provides sample parameters that could be used for realization of the adiabatic gates. Identical circuit parameters are assumed for the $i=1..k$ control(target) qubits, different from the ones of the sole target(control) qubit (labeled by $0$) for the Toffoli(fan-out) gate. The parameters are tuned  to establish degeneracy between control-target pairs $\delta_{2_t,2_c}^{3_t1_c}=0$ (see Fig.~\ref{FigSupCon}(b) in the main text), and the dynamics is controlled by the two degenerate  inter-component (i.e. control-target) $B_{1}=\frac{\mathbb{C}_{0i}}{\sqrt{Z_0Z_i}}$ and intra-component $B_{2}=\frac{\mathbb{C}_{ij}}{\sqrt{Z_iZ_j}}$ exchange interactions. While $B_{1}$ is required for the formation and following of the dark state, $B_{2}$ causes bit flip errors. Table III shows that reducing the ratio of coupling capacitance  over qubit capacitance  $\frac{C_{xi}}{C_i}$ suppresses $\frac{B_2}{B_1}$. Hence, the  exchange Hamiltonian in the rotating frame approximation may be limited to the desired pair for the dark-state formation i.e. $B_{1}|2_t2_c\rangle\langle3_t1_c|$. 

\begin{center}
\begin{tabular}{ |c c c c c |}
 \hline
  No. &  C$_{xi}/$C$_{i}$ &\quad  C$_0$ ${\scriptstyle (pF)}$\quad &    $B_{1}$${\scriptstyle (2\pi \text{MHz})}$  & \quad $B_{2}/B_1$ 	  \\
 \hline
  Toffoli &  &      &     &  \\
 \hline
1  & \quad $10^{-1}$\qquad  &\quad  24.54\quad& 140 & \quad 2.7$\times10^{-3}$  \\
2 & \quad $10^{-2}$ \qquad &\quad 22.79 \quad& 16 & \quad 2.7$\times10^{-4}$  \\
3 &\quad  $10^{-3}$ \qquad &\quad 22.62 \quad& 1.6 & \quad 2.7$\times10^{-5}$  \\
  \hline
   Fan-out   & &      &  &  \\
  \hline
  1  & \quad  $10^{-2}$ \qquad &\quad 1.40 \quad& 123 & \quad 8.5$\times10^{-3}$ \\
2 & \quad $10^{-3}$ \qquad &\quad1.48 \quad& 12.6 & \quad 8.5$\times10^{-4}$ \\
3 & \quad $10^{-4}$ \qquad &\quad 1.48 \quad& 1.3 &\quad 8.5$\times10^{-5}$  \\
  \hline \end{tabular}
\captionof{table}{ Example parameters (rounded values) for the implementation of adiabatic multi-qubit gates in superconducting circuits. The  scheme parameters in this table for  Toffoli (fan-out) are  E$_0=20$(1.25)ns$^{-1}$, E$_i$=1.25(20)ns$^{-1}$, C$_i=$5(20)pF. The  detuning parameters in Toffoli (fan-out) are $\delta_{2_t0_c}^{1_t1_c}/2\pi$=3.2(18.6)GHz, $\delta_{0_t1_c}^{1_t0_c}/2\pi$=2.2(5.4)GHz, $\alpha_t/2\pi$=0.87(4)GHz and $\alpha_{c}/2\pi$=4(13.4)GHz.
 The presented coupling and detuning strengths are calculated for $k=2$. Adjustment is required for other qubit numbers $k$ to preserve the same coupling parameters. For example in k=20 Toffoli (fan-out) gate, the adjusted value of row No.2 is $C_0=$21.8953(1.3854)pF.}
\end{center}

 \begin{figure}
\centering
       \subfloat{%
\includegraphics[width=.4\textwidth]{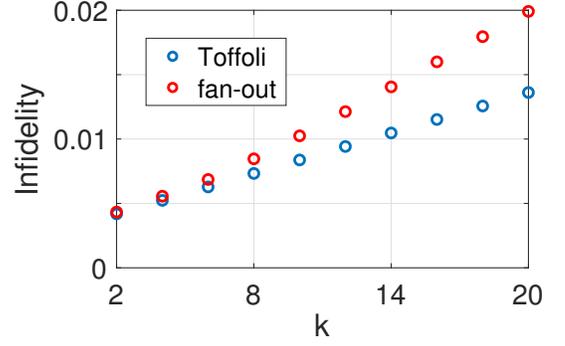}}\hfill
\caption{Infidelity of the adiabatic superconducting multi-qubit gate as a function of the number of qubits $k$. The fidelity of the two gates are evaluated for the parameters of row No. 2 in table III, the damping rate $\gamma/2\pi=5.7$kHz,  and the driving strengths, $\Omega_c/2\pi=$48(160)MHz,  $\Omega_t/2B_1$=1/4 (1.0) for the Toffoli(fan-out) gates.
}\label{FigFidSupCon}
\end{figure}

\subsection*{Gate errors}

Here we  address the effects of dissipation, errors in population rotations, excitation exchange and non-adiabatic dynamics on the gate fidelity.
The dissipation results to $\sqrt{\gamma}|0 \rangle \langle1|$ and $\sqrt{2\gamma}|1 \rangle \langle 2 |$ Lindblad terms associated with decay and de-phasing in qubit dynamics. To estimate the errors for gates with large qubit numbers, we only consider the decay phenomenologically. The average dissipation error in the Toffoli and fan-out gate is estimated by
 \begin{eqnarray}
 \label{EqEdiss}
&&E^{t}_{dis}= \frac{k}{2}\frac{2\pi}{\Omega_c}\gamma+\frac{1}{2}(\frac{4\pi}{\Omega_c}+\frac{2\pi}{\Omega_t})\gamma \\ \nonumber
&&E^{f}_{dis}= \frac{1}{2}  \frac{2\pi}{\Omega_c}\gamma+ \frac{k}{2} (\frac{4\pi}{\Omega_c}+\frac{2\pi}{\Omega_t})\gamma
\end{eqnarray}
where the first and second terms address the errors in control and target units respectively.
The level anharmonicity is necessary to drive specific transition, and the rotation errors,  discussed also in atomic systems, are given as
\begin{eqnarray}
&&E_{rot}^{t}=k\frac{\Omega_c^2}{\alpha_c^2}+\frac{\Omega_t^2}{\alpha_t^2}\\ \nonumber
&&E_{rot}^{f}=\frac{\Omega_c^2}{\alpha_c^2}+k\frac{\Omega_t^2}{\alpha_t^2}
\end{eqnarray}
The intra-component resonant exchange interaction among target or control qubits leads to an average error of
 \begin{eqnarray}
 \label{EqExEr1}
&&E_{ex1}^{t}= \frac{2}{2^{k}} \sum\limits_{j=0}^{k/2} \left(\begin{array}{c} k\\ j \end{array}\right) (\frac{2\pi }{\Omega_c})^2 j B_2^2\\  \nonumber
&&E_{ex1}^{f}= \frac{2}{2^{k}} \sum\limits_{j=0}^{k/2} \left(\begin{array}{c} k\\ j \end{array}\right) (\frac{4\pi }{\Omega_c}+\frac{2\pi }{\Omega_t})^2 jB_2^2
\end{eqnarray}
and the off-resonant inter-component exchange interactions yield an error of
 \begin{equation}
 \label{EqExEr2}
E_{ex2}=  \frac{1}{2}[(\frac{B_1}{\delta_{2_t0_c}^{1_t1_c}})^2+(\frac{B_1}{\delta_{1_t0_c}^{0_t1_c}})^2].
\end{equation}
Considering the similarity of the dark states level scheme in figures \ref{FigDarkState}a and \ref{FigSupCon}c,d, the average error due to non-adiabatic transitions for a Gaussian excitation pulse is estimated by Eq. (7) for both the Toffoli and fan-out gates.

The resulting accumulated gate error, i.e., the sum of all these terms, is plotted as a function of the number of qubits $k$ in Fig.~\ref{FigFidSupCon} for the parameters presented in row No. 2 of table III. In Fig.~\ref{FigFidSupCon}, the decay rate of  $\gamma/2\pi=5.7$kHz is chosen, corresponding to the relaxation time of $T_1=30\mu s$, which are realistic values, cf., $T_1=70\mu s$ and $T_2=95\mu s$ excitation and coherence times reported in \cite{Rig12}.
With realistic parameters, we achieve infidelities below $0.02$ in superconducting circuits for $k<20$.


\begin{thebibliography}{100}
\bibitem{FigCap}The main infidelity of the dark state evolution in Fig.~\ref{FigDarkStateEvol} is due to non-adiabatic errors and the  $B_2$ interaction terms. The adiabatic errors are minimum at the beginning, middle and end of the Gaussian pulse where $\dot{\Omega}=0$. Effects of $B_2$ are present in the first and third two-photon excitation in Toffoli and Fan-Out gates, respectively, making the effect benign in Fan-out gate due to the minor population while causing a visible deviation in the Toffoli gate.
\bibitem{Point2}  In the analytical  error estimates in Sec. IV, we used the maximum interaction strength, i.e. the interaction of neighboring sites. This is because the averaged gate error is  mainly affected  by the qubit configurations with large intra-component interactions. This conservative estimate yields better agreement with lattice simulations than, e.g., the distance averaged interaction strength.
\bibitem{Point3}The polarizability is not the same for different Rydberg levels, and the typical Rydberg scaling
laws are not applicable in the presence of external fields. We may, however, summarize the main
effect of, e.g., reducing the principal quantum numbers: For lower $n$-values,
a stronger electric field is required to tune the appropriate Rydberg pairs into resonance.
Higher decay rates specially affect the Toffoli gate due to its higher time integrated population of the Rydberg excited state, while
the second rotation error decreases, and higher optical transition dipole moments and hence achievable Rabi frequency permit faster near-adiabatic operation. Finally, the critical distance explained in Appendix A.4 would be smaller to account for the weaker interaction coefficients.
After optimization of the physical parameters, we hence expect similar performance with moderate changes of the principal numbers.
\end{thebibliography}
\end{document}